\documentclass[a4paper,fleqn,usenatbib]{mnras}

\usepackage{newtxtext,newtxmath}

\usepackage[T1]{fontenc}
\usepackage{ae,aecompl}

\usepackage{graphicx}	
\usepackage{amsmath}	
\usepackage{amssymb}	
\usepackage[english]{babel}
\usepackage{subfig}
\usepackage{tikz}
\usepackage{float}
\usepackage{natbib}
\usepackage{apalike}
\usepackage{wasysym}
\usepackage{ragged2e}

\title[GRB hosts from chemical abundances]{The nature of GRB host galaxies from chemical abundances}

\author[Palla et al.]{
Marco Palla,$^{1}$\thanks{E-mail: marco.palla@phd.units.it}
Francesca Matteucci$^{1,2,3}$,
Francesco Calura$^{4}$,
Francesco Longo$^{3,5}$
\\
$^{1}$ Dipartimento di Fisica, Sezione di Astronomia, Universit{\'a} degli Studi di Trieste, via G. B. Tiepolo 11, I-34131, Trieste, Italy\\
$^{2}$ INAF, Osservatorio Astronomico di Trieste, via G. B. Tiepolo 11, I-34131, Trieste, Italy\\
$^{3}$ INFN, Sezione di Trieste, via A. Valerio 2, I-34100, Trieste, Italy\\
$^{4}$ INAF, Osservatorio Astronomico di Bologna, via P. Gobetti 93/3, I-40129, Bologna, Italy\\
$^{5}$ Dipartimento di Fisica, Universit{\'a} degli Studi di Trieste, via A. Valerio 2, I-34100, Trieste, Italy
}

\date{Accepted XXX. Received YYY; in original form ZZZ}

\pubyear{2019}

\begin{document}
\label{firstpage}
\pagerange{\pageref{firstpage}--\pageref{lastpage}}
\maketitle

\begin{abstract}
We try to identify the nature of high redshift long Gamma-Ray Bursts (LGRBs) host galaxies by comparing the observed abundance ratios in the interstellar medium with detailed chemical evolution models accounting for the presence of dust.  We compared measured abundance data from LGRB afterglow spectra to abundance patterns as predicted by our models for different galaxy types. We analysed in particular $[X/Fe]$ abundance ratios (where $X$ is $C$, $N$, $O$, $Mg$, $Si$, $S$, $Ni$, $Zn$) as functions of $[Fe/H]$. Different galaxies (irregulars, spirals, ellipticals) are, in fact, characterised by different star formation histories, which produce different $[X/Fe]$ ratios ($\lq\lq$time-delay model"). This allows us to identify the morphology of the hosts and to infer their age (i.e. the time elapsed from the beginning of star formation) at the time of the GRB events, as well as other important parameters. Relative to previous works, we use newer models in which we adopt updated stellar yields and prescriptions for dust production, accretion and destruction. We have considered a sample of seven LGRB host galaxies. Our results have suggested that two of them (GRB 050820, GRB 120815A) are ellipticals, two (GRB 081008, GRB 161023A) are spirals and three (GRB 050730, GRB 090926A, GRB 120327A) are irregulars. We also found that in some cases changing the initial mass function can give better agreement with the observed data. The calculated ages of the host galaxies span from the order of $10$ $Myr$ to little more than $1$ $Gyr$.
\end{abstract}





\section{Introduction}
\label{s:intro}
Gamma-Ray Bursts (GRBs) are sudden extremely powerful flashes of gamma radiation. They originate at cosmological distances and last from timescales of millisecond to $10^3$ seconds. Thanks to \citet{Kouveliotou93} classification, GRBs with a duration longer than $2$ seconds are called long GRBs (LGRBs). This GRB class is associated with the death of very massive stars, thanks to the study of GRB afterglows, i.e. the fading follow-up of the prompt emission, occurring at longer wavelength (X-ray, optical, radio). As a matter of fact, from the lightcurves and spectra analyses of these afterglows, a firm association with a core collapse supernova (CC-SN) was found for 27 LGRBs, until 2016 (\citealt{Hjorth16}).\\
Since LGRBs are likely to be related to the death of massive stars, and for the fact that these stars have very short lives, probing the interstellar medium (ISM) that surrounds the GRB site means probing the ISM of a star-forming region. In this scenario, with the constantly increasing number of high redshift GRBs analysed so far, afterglow spectra can be used to probe star-forming galaxies, in particular those at high redshift. 
Furthermore, understanding of the nature of GRB host galaxies can give stringent constraints on GRB progenitor models, favouring the single progenitor (collapsar by \citealt{MacFadyen99}; \citealt{Woosley06} or millisecond magnetar by \citealt{Wheeler00}; \citealt{Bucciantini09}) or the binary progenitor models (e.g. \citealt{Fryer05}; \citealt{Demeters08}; \citealt{Podsiadlowski10}). Many attempts have been made in the past to characterise GRB host galaxies (e.g. \citealt{LeFloch03}; \citealt{Fruchter06}; \citealt{Savaglio09}; \citealt{Levesque10}; \citealt{Boissier13}; \citealt{Schulze15}; \citealt{Perley16b}; \citealt{Arabsalmani18}) and it is still debated if LGRB hosts sample the general star-forming galaxy population or if they represent a distinct galaxy population.\nocite{Bignone17}\\[0.1cm]
Following the idea developed by \citet{Calura09} and adopted also by \citet{Grieco14}, in this paper we use chemical evolution models for different galaxy morphological types (irregular, spiral, elliptical) which predict the abundances of the main chemical elements ($H$, $He$, $C$, $N$,$\alpha$-elements\footnote{elements synthetised by capture of $\alpha$ particles. Examples are $O$, $Mg$, $Si$, $S$.}, $Fe$, $Ni$, $Zn$, etc.), to identify the nature of GRB host galaxies. The basic idea beneath this procedure derives from the $\lq\lq$time-delay model" (\citealt{Matteucci01}, \citeyear{Matteucci12}), which explains the observed behaviour of $[X/Fe]$\footnote{by definition: $[X/Y] = \log(X/Y) - \log(X_\odot/Y_\odot)$, where $X$, $Y$ are abundances in mass in the ISM for the object studied and $X_\odot$, $Y_\odot$ are solar abundances in mass.} vs. $[Fe/H]$ , with $X$ being any chemical element, as due to the different roles played by core collapse and type Ia SNe (white dwarfs exploding in binary systems) in the galactic chemical enrichment. Based on the fact that $\alpha$-elements to $Fe$ ratio evolution is predicted to be quite different in different star formation (SF) regimes (\citealt{Matteucci90}; \citealt{Matteucci01}), this model foresees for different morphological types a well-defined different behaviour of the $[\alpha/Fe]$ vs. $[Fe/H]$ abundance diagrams.\\
The models we are adopting for diffent galaxy types differ by the star formation history and take into account possible condensation of the main metals ($C$, $\alpha$-elements, $Fe$, $Ni$) into dust. Observations of LGRBs in mid-IR and radio bands, indeed, show clearly the presence of dusty environments in many hosts (e.g. \citealt{Perley09}, \citeyear{Perley13}, \citeyear{Perley17}; \citealt{Greiner11}; \citealt{Hatsukade12}; \citealt{Hunt14}). This latter fact, coupled with the undetectability of dust grains by optical/UV spectroscopic measurements, is fundamental for our aim of understanding the nature of the hosts using abundance patterns: without any consideration of dust presence, our interpretation of observational data would risk to bring to misleading conclusions. With respect to the previous works of \citet{Calura09} and \citet{Grieco14}, based on the chemical evolution models with dust by \citet{Calura08}, in this paper we adopt improved chemical evolution models with dust. We use newer and more accurate prescriptions for dust production (from \citealt{Piovan11} and \citealt{Gioannini17a}) and other dust processes in the ISM (from \citealt{Asano13}), as well as for the stellar yields (from \citealt{Karakas10}; \citealt{Doherty14a}, \citeyear{Doherty14b}; \citealt{Nomoto13}). With respect to the considered host galaxies, we take afterglow spectra already studied by \citet{Calura09} and \citet{Grieco14}, plus a couple of systems never considered before such analysis.\\ 
The paper is organised as follows: Section \ref{s:sample} shows and briefly explains the observational data adopted. Section \ref{s:chem_model} describes the chemical evolution models adopted, specifying also the dust prescriptions (production, accretion and destruction) used throughout this work. In Section \ref{s:results} we explain the parameter values in the adopted models and we show the results derived from the comparison between the prediction given by the models and the abundance data for the analysed hosts. Finally, in Section \ref{s:conclusion} some conclusions are drawn. 

\section{Host galaxies sample}
\label{s:sample}
In order to constrain the nature of GRB host galaxies we have chosen from the literature bursts with a quite large number of observed abundances from the environment: GRB 050730, GRB 050820 (\citealt{Prochaska07}), GRB 081008 (\citealt{DElia11}), GRB 090926A (\citealt{DElia10}), GRB 120327A (\citealt{DElia14}), GRB 120815A (\citealt{Kruhler13}), GRB 161023A (\citealt{deUgarte18}). In Table \ref{t:observation} the observational data (redshift, abundance ratios) are shown for each host studied in our analysis.\\
We decided to include in our sample data already used in the previous GRB hosts identification works of \citet{Calura09} (GRB 050730, GRB 050820) and \citet{Grieco14} (GRB 081008, GRB 120327A, GRB 120815A). The main reason of their inclusion is to test the results obtained with older chemical evolution models (\citealt{Calura08}) containing less updated stellar yields and dust prescriptions (production by stars, accretion, destruction). In this way, we can see if newer models lead to different conclusions with respect to older ones, highlighting the importance of using more accurate models to reach more robust conclusions.\\
We note finally that in Table \ref{t:observation} we do not present $[C/Fe]$ and $[O/Fe]$ ratios, that are available in almost all the studies considered. This decision was made because they are all lower/upper limits that do not give additional information to what predicted by other elements, or abundances affected by biases (lines saturation, blending) in their determination.

\begin{table*} 
\caption{Gamma-Ray Burst, redshift and abundance ratios associated to the host galaxies object of the study (references in Section \ref{s:sample}). All abundance ratios are normalized to \citet{Asplund09} solar abundances.}
\begin{tabular}{c | *{7}c}
\hline
 & GRB 050730 & GRB 050820 & GRB 081008 & GRB 090926A & GRB 120327A & GRB 120815A & GRB 161023A \\
\hline
$z$ & $3.969$ & $2.165$ & $1.968$ & $2.107$ & $2.815$ & $2.360$ & $2.710$\\
\hline
$[Fe/H]$ & $-2.59\pm0.10$ & $-1.69\pm0.10$ & $-1.19\pm0.11$ & $-2.29\pm0.09$ & $-1.73\pm0.10$ & $-2.18\pm0.11$ & $1.81\pm0.04$\\
$[N/Fe]$ & $-0.47\pm0.14$ & $>0.44$ & - & $-0.97\pm0.08$ & $0.28\pm0.15$ & - & -  \\ 
$[Mg/Fe]$ & $<0.88$ & $0.91\pm0.14$ & - & $>-0.83\pm0.12$ & $0.46\pm0.14$ & - & $0.32\pm0.20$\\
$[Si/Fe]$ & $>-0.23$ & $>0.50$ & $0.32\pm0.15$ & $-0.12\pm0.12$ & $0.61\pm0.15$ & $\apprge1.02\pm0.22$ & $0.37\pm0.06$\\
$[S/Fe]$ & $0.35\pm0.14$ & $1.08\pm0.14$ & - & $0.34\pm0.13$ & $0.34\pm0.13$ & $\apprle 1.31\pm0.28$ & $0.66\pm0.06$\\
$[Ni/Fe]$ & $-0.06\pm0.14$ & $0.16\pm0.14$ & $-0.10\pm0.16$ & $0.28\pm0.16$ & $0.10\pm0.13$ & $0.22\pm0.16$ & - \\
$[Zn/Fe]$ & - & $1.04\pm0.14$ & $0.67\pm0.15$ & - & $0.56\pm0.15$ & $1.10\pm0.15$ & $0.70\pm0.08$\\
\hline
\end{tabular}
\label{t:observation}
\end{table*}

\section{Chemical evolution models including dust}
\label{s:chem_model}
We trace the evolution of chemical abundances in galaxies of different morphological types by means of chemical evolution models including dust evolution. These models relax the so called instantaneous recycling approximation (IRA), taking into account stellar lifetimes. All the models assume that galaxies form by primordial gas infall which accumulate into a preexisting dark matter halo.\\
A fundamental parameter for these models is the birthrate function $B(m,t)$, which represents the number of stars formed in the mass interval $[m, m + dm]$ and in the time interval $[t, t + dt]$. It is expressed as the product of two independent functions, in this way:
\begin{equation}
   B(m, t) = \psi(t)\phi(m),
\end{equation}
where the term $\psi(t)$ is the star formation rate (SFR), whereas $\phi(m)$ represents the initial mass function (IMF).\\
The SFR is the rate at which stars form per unit time and it is generally expressed in units of $M_\odot yr^{-1}$. To parametrise the SFR, in our models we adopt the Schmidt-Kennicutt law (\citealt{Schmidt59}; \citealt{Kennicutt89}):
\begin{equation}
    \psi(t) = \nu G(t)^k.
\end{equation}
In this expression, $\nu$ is the star formation efficiency, namely the inverse of the time scale of star formation (expressed in $Gyr^{-1}$), which varies depending on the morphological type of the galaxy (see Table \ref{t:models}). In particular, the variation of $\nu$ determines the different SFR in galaxies of different morphological type, decreasing from ellipticals to spirals and irregulars. $G(t) = M_{ISM} (t)/M_{inf}$ is the ISM mass fraction relative to the infall mass, i.e. the total mass accumulated until the final evolutionary time $t_f$, which is set to be $14$ $Gyr$ for all the models. The parameter $k$ is set equal to $1$.\\
The IMF represents the mass distribution of stars at their birth. It is assumed to be constant in space and time and normalised to unity in the mass interval $[0.1M_\odot , 100M_\odot ]$. In our work, the calculation for all galaxies are performed first using a \citet{Salpeter55} IMF:
\begin{equation}
    \phi_{Salp}(m)=0.17 m^{-(1+1.35)}.
\end{equation}
For elliptical galaxies, the computations are done also with a top-heavy single-slope IMF,
\begin{equation}
    \phi_{top}(m)=0.16 m^{-(1+1.1)},
\end{equation}
since in more massive ellipticals an overabundance of massive stars at early times is necessary to explain their observational data (e.g. \citealt{Gibson97}; \citealt{Weidner13}).\\
For the spirals and irregulars instead, in addition to the \citet{Salpeter55}, we use also a \citet{Scalo86} IMF, derived for the solar vicinity:
\begin{gather}
\phi_{Scalo}(m) = \bigg \{\begin{array}{rl}
					0.19\cdot m^{-(1+1.35)} & m\leq 2 M_\odot \\
					0.24\cdot m^{-(1+1.7)} \hspace*{0.1cm}    & m > 2 M_\odot,\\
				  \end{array}
\end{gather}
which fits better the features of spiral disks than the \citet{Salpeter55} (\citealt{Chiappini01}; \citealt{Romano05}).

\subsection{Chemical evolution equations}
\label{ss:chem_eq}
The basis of every chemical evolution work are the chemical evolution equations for the various chemical elements. For a given element $i$, they have the following form:
\begin{equation}
\dot{G}_i(t)=-\psi(t)X_i(t)+ R_i(t) + \dot{G}_{i,inf}(t) - \dot{G}_{i,w}(t),
\label{e:chem_evo}
\end{equation}
where $G_i(t)=G(t)X_i(t)$ is the mass of the element $i$ in the ISM normalised to the infall mass and $X_i(t)$ represents the fraction of the element $i$ in the ISM at a certain time $t$.\\
The four terms on the right side are the following:
\begin{enumerate}
    \item $-\psi(t)X_i (t)$ represents the rate at which the element $i$ is removed from the ISM due to the star formation process.
    \item $R_i (t)$ is the rate at which the element $i$ is restored into the ISM from stars thanks to SN explosions and stellar winds. Inside this term the nucleosynthesis prescriptions of the specific element $i$ are taken into account (see \ref{sss:nucleosynthesis}). In order to relax the IRA, $R_i (t)$ has the following form, as shown by \citet{MatteucciGreggio86}:
\begin{multline}
R_i(t)=\int_{M_L}^{M_{Bm}}\!{\psi(t-\tau_m)Q_{mi}(t-\tau_m)\phi(m)}\,dm +\\
+A\int_{M_{Bm}}^{M_{BM}}\phi(m) \bigg[ \int_{\mu_{min}}^{0.5} f(\mu)	\psi(t-\tau_{m2})Q_{mi}(t-\tau_{m2})\,d\mu \bigg] \,dm +\\
+ (1-A)\int_{M_{Bm}}^{M_{BM}}\psi(t-\tau_m)Q_{mi}(t-\tau_m)\phi(m)\,dm +\\ +\int_{M_{BM}}^{M_{U}}\psi(t-\tau_m)Q_{mi}(t-\tau_m)\phi(m)\,dm . \\	
\label{e:restitution_stars}
\end{multline}
The first integral is the rate at which an element $i$ is restored into the ISM by single stars with masses in the range $[M_L,M_{B_m}]$, where
$M_L$ is the minimum mass at a certain time $t$ contributing to chemical enrichment (for $t_f=14Gyr$, $M_L=0.8M_\odot$ ) and
$M_{B_m}$ is  the  minimum mass for a binary system to give rise to a type Ia SN ($M_{B_m}=3M_\odot$). The quantities $Q_{mi}(t-\tau_m)$, where $\tau_m$ is the lifetime of a star of mass $m$, contain all the information about stellar nucleosynthesis for elements either produced or destroyed inside stars or both (\citealt{Talbot71}).\\
The second term represents the material restored by binaries, with masses between $M_{B_m}$ and $M_{B_M}=16M_\odot$, which have the right properties to explode as type Ia SNe. For these SNe a single degenerate scenario (SD) is assumed, where a $C$-$O$ white dwarf explodes after it exceeds the Chandrasekar mass ($1.44M_\odot$). $A$ is the parameter representing the fraction of binary systems able to produce a type Ia SN and its value is set to reproduce the observed rate of type Ia SNe. In this term both $\psi$ and $Q_{mi}$ refer to the time $t-\tau_{m2}$ where $\tau_{m2}$ indicates the lifetime of the secondary star of the binary system, which regulates the explosion timescale. $\mu=m_2/m_B$ is the ratio between the mass of the secondary component ($m_2$) and the
total mass of the binary ($m_B$) and $f(\mu)$ represent the distribution of this ratio.\\
The third integral represents the contribution given by single stars lying in the mass range [$M_{B_m} , M_{B_M}]$ which do not produce type Ia SNe events. If the mass $m > 8M_\odot$ , they explode as CC-SNe.\\
The last term of \eqref{e:restitution_stars} refers to the material recycled back to the ISM by stars more massive than $M_{B_M}$, i.e. by the high mass CC-SNe up to $M_U=100M_\odot$.
    \item $\dot{G}_{i,inf}(t)$ represents the rate of infall of gas of the $i$-th element in the system. It is expressed in this way:
    \begin{equation}
\dot{G}_{i,inf}(t)= \frac{\Gamma}{M_{tot}(t_f)} X_{i,inf} e^{-t/\tau_{inf}},
\label{e:infall_rate}
\end{equation}
where $\Gamma$ is the normalisation constant, constrained to reproduce the infall mass at the final time $t_f$, and $X_{i,inf}$ the fraction of the element $i$ in the infalling gas, which has primordial composition. $\tau_{inf}$ is the infall timescale, defined as the characteristic time at which half of the total mass of the galaxy has assembled and is set to satisfy observational constraints for the studied galaxies. This is the other parameter which varies $\lq\lq$progressively" with galactic type, increasing from elliptical to spirals and irregulars.
    \item The last term of Equation \eqref{e:chem_evo} represents the outflow rate of the element $i$ due to galactic winds (GWs), developing when the thermal energy of the gas (heated by SNe explosions) exceeds its binding energy. The outflow rate has this form:
    \begin{equation}
\dot{G}_{i,w}(t)= k_i \psi(t),
\end{equation}
where $k_i$ is the wind rate parameter for the element $i$, a free parameter chosen in order to reproduce the galaxy features. In our models we do not use differential winds, so $k_i$ will be the same for all elements.
\end{enumerate}

\subsubsection{Nucleosynthesis prescriptions\\}
\label{sss:nucleosynthesis}
We compute in detail the contribution to chemical enrichment of the ISM of low-intermediate mass stars (LIMS), type Ia and CC-SNe. To do this we adopted specific stellar yields for all these stars. The yields are the amount of both newly formed and pre-existing elements injected into the ISM by dying stars.\\
In this paper we adopt mass and metallicity dependent stellar yields:
\begin{enumerate}
    \item for LIMS ($0.8M_\odot<m<9M_\odot$) we use yields by \citet{Karakas10} for stars with mass lower than $6M_\odot$, while for super-AGB (SAGB) stars and e-capture SNe, with mass between $6M_\odot$ and $9M_\odot$ , we use yields by  \citeauthor{Doherty14a}(\citeyear{Doherty14a}, \citeyear{Doherty14b}).
    \item For massive stars, that explode as CC-SNe ($m>9M_\odot$), we adopt yields by \citet{Nomoto13}. For nitrogen, calculation are performed also with the prescriptions by \citet{Matteucci86}.
    \item For type Ia SNe we use the yields by \citet{Iwamoto99}.
\end{enumerate}

\subsection{Dust evolution equation}
\label{ss:dust_eq}
Adopting the same formalism used in previous works on chemical evolution models with dust (\citealt{Dwek98}; \citealt{Calura08}; \citealt{Gioannini17a}), the equation governing the dust evolution is quite similar to \eqref{e:chem_evo}, but it includes other terms describing dust processes in the ISM. For a given element $i$, we have:
\begin{equation}
\begin{split}
\dot{G}_{i,dust}= - \psi(t) X_{i,dust}(t) + R_{i,dust}(t) + \dot{G}_{i,dust,accr}(t) +\\
-\dot{G}_{i,dust,destr}(t) - \dot{G}_{i,dust,w}(t), \hspace*{1.5cm}
\end{split}
\label{e:dust_evo}
\end{equation}
where $G_{i,dust}(t)=G(t) X_{i,dust}(t)$ is the mass of an element $i$ in the dust phase normalised to the infall mass and $X_{i,dust}(t)$ is the fraction of the element $i$ in the dust phase at a certain time $t$.\\
The five terms on the right side of Equation \eqref{e:dust_evo} are the following:
\begin{enumerate}
    \item The first term concerns the rate of dust astration. In other words, this is the process of removal of dust from the ISM due to star formation.
    \item $R_{i,dust}(t)$, similarly to $R_i(t)$ for Equation \eqref{e:chem_evo}, is the rate at which the element $i$ in the dust phase is restored into the ISM. The term is also called dust production rate (DPR).
    \item The third term is the dust accretion rate (DAR) for the element $i$, which is the rate of dust mass enhancement due to grain growth by accretion processes in the ISM.
    \item $\dot{G}_{i,dust,destr}(t)$ is the dust destruction rate (DDR) for the $i$-th element, namely the rate of dust mass decrease by grain destruction.
    \item The last term of Equation \eqref{e:dust_evo} indicates the rate of dust, in the form of element $i$, expelled by GWs. In the model we assume that dust and gas in the ISM are coupled, so the wind parameters are the same for the elements in gas and dust.
\end{enumerate}
In the next paragraphs we will discuss the second, third and fourth terms in more detail.

\subsubsection{Dust production}
\label{ss:d_prod}
The interstellar dust is first produced by stars: depending
on the physical structure of the progenitor (type of star, mass, metallicity), different amounts of dust species can originate.\\
We can summarise the second term of Equation \eqref{e:dust_evo} in this way (the complete expression can be found in \citealt{Gioannini17a}):
\begin{equation}
    R_{i,dust}(t)=\delta_i^{AGB}R_i^{LIMS}(t)+\delta_i^{CC}R_i^{CC-SN}(t).
\end{equation}
In other words, we have the same expression of \eqref{e:restitution_stars} without considering type Ia SNe contribution and with the addition of the terms $\delta_i^{AGB}$, $\delta_i^{CC}$. These terms are the condensation efficiencies and represent the fraction of the element $i$ expelled by stars (AGB and CC-SNe, respectively) which goes into the ISM in the dust phase.\\
Following \citet{Gioannini17a}, the dust sources considered in this work are:
\begin{enumerate}
    \item AGB (LIMS): in LIMS, the cold envelope during the AGB phase is the best environment in which nucleation and formation of dust seeds can occur, since previous phases do not present favourable conditions (small amount of ejected material, wind physical conditions) for producing dust. In the dust production process, stellar mass and metallicity play a key role in determining the dust species formed: this happens because $m$ and $Z$ are crucial to set the number of thermal pulses occurring, which define the surface composition of the star (e.g. \citealt{Ferrarotti06}; \citealt{DellAgli17}).\\
    In this paper we adopt the condensation efficiencies, dependent both on mass and metallicity, computed by \citet{Piovan11}, already presented and used in \citet{Gioannini17a}.
    \item CC-SNe: this is the other fundamental source of dust besides AGB stars. Evidence of dust presence in historical supernova remnants, such as SN1987A (e.g. \citealt{Danziger91}) Cas A, Crab Nebula, were observed (\citealt{Gomez13} and references therein). In particular from SN1987A observations, we now know that this SN produced up to $0.7M_\odot$ of dust. Despite of this, the picture is far from being totally clear. This is due to the lack in understanding the amount of dust destroyed by the reverse shock of the explosion after the initial production (see \citealt{Gioannini17a} for a more detailed discussion).\\
    Also in this case we adopt the condensation efficiencies provided by \citet{Piovan11}, presented and used in \citet{Gioannini17a}. These $\delta_i^{CC}$ take into account both the processes of dust production and destruction by CC-SNe, but most importantly give us the possibility to choose between three different scenarios for the surrounding environment: low density ($n_H=0.1$ $cm^{-3}$), intermediate density ($n_H=1$ $cm^{-3}$) and high density ($n_H=10$ $cm^{-3}$). The higher is the density, the higher is the resistance that the shock will encounter, and the higher will be the dust destroyed by this shock. Between the three possibilities, in this work we adopt only $\delta_i^{CC}$ for $n_H=0.1$ $cm^{-3}$ and $n_H=1$ $cm^{-3}$. We make this choice looking at \cite{Gioannini17b}, where the intermediate density scenario quite well reproduce the amount of dust detected in some high redshift ellipticals and the dust-to-gas ratio (DGR) in spirals of the KINGFISH survey (\citealt{Kennicutt11}), whereas low density $\delta_i^{CC}$ are more indicated to explain DGR observed in the Dwarf Galaxy Survey (\citealt{Madden13}).
\end{enumerate}
In this work, we assume that type Ia SNe do not produce any dust, following what done in the works of \citeauthor{Gioannini17a} (\citeyear{Gioannini17a}, \citeyear{Gioannini17b}). As a matter of fact, both from the observational (e.g. \citealt{Gomez12}) and the theoretical (\citealt{Nozawa11}) point of view there are no evidences that these SNe produce a significant dust amount (see \citealt{Gioannini17a} for more information).
\begin{table*} 
\caption{Input parameters for the chemical evolution models adopted in this work. The parameters of the reference models for the three different morphological scenarios (irregular galaxy, spiral disk, elliptical galaxy) are in bold.}
\begin{tabular}{c | *{6}c}
\hline
Model & $M_{inf} [M_\odot]$ & $\tau_{inf} [Gyr]$ & $\nu [Gyr^{-1}]$ & $K_i$ & IMF & $\delta^{CC}$\\
\hline
\textbf{I}, I$-$ & \boldmath$5\cdot10^9$ & \boldmath$10$ & \boldmath$0.1$ & \boldmath$0.5$ & \textbf{\citeauthor{Salpeter55}} & \boldmath$\delta_{HP}$\unboldmath, $\delta_{MP}$ \\
I$l$, I$l-$ & $5\cdot10^8$ & $10$ & $0.02$ & $1$ & \citeauthor{Salpeter55} & $\delta_{HP}$, $\delta_{MP}$\\
IS, IS$-$ & $5\cdot10^9$ & $10$ & $0.1$ & $0.5$ & \citeauthor{Scalo86} & $\delta_{HP}$, $\delta_{MP}$\\
IS$l$, IS$l-$ & $5\cdot10^8$ & $10$ & $0.02$ & $1$ & \citeauthor{Scalo86} & $\delta_{HP}$, $\delta_{MP}$\\
\hline
\textbf{Sp}, Sp$+$ & \boldmath$5\cdot10^{10}$ & \boldmath$7$ & \boldmath$1$ & \boldmath$0.2$ & \textbf{\citeauthor{Salpeter55}} & \boldmath$\delta_{MP}$\unboldmath, $\delta_{HP}$\\
SpS, SpS$+$ & $5\cdot10^{10}$ & $7$ & $1$ & $0.2$ & \citeauthor{Scalo86} & $\delta_{MP}$, $\delta_{HP}$\\
\hline
\textbf{E}, E$+$ & \boldmath$10^{11}$ & \boldmath$0.3$ & \boldmath$15$ & \boldmath$10$ & \textbf{\citeauthor{Salpeter55}} & \boldmath$\delta_{MP}$\unboldmath, $\delta_{HP}$\\
E$m$, E$m+$ & $10^{12}$ & $0.2$ & $25$ & $20$ & \citeauthor{Salpeter55} & $\delta_{MP}$, $\delta_{HP}$ \\
ET$m$, ET$m+$ & $10^{12}$ & $0.2$ & $25$ & $20$ & Top-heavy & $\delta_{MP}$, $\delta_{HP}$\\
\hline
\end{tabular}
\label{t:models}
\end{table*}

\subsubsection{Dust accretion}
\label{sss:d_accretion}
During galactic evolution, dust grains in the ISM can grow in size due to accretion by metal gas particles on the surface of these grains. This process, occurring mostly in the coldest and densest regions of the ISM, i.e. molecular clouds, has the power to increase the global amount of interstellar dust. For this reason, dust accretion is a fundamental ingredient in dust chemical evolution, as pointed out by many studies (e.g. \citealt{Dwek98}; \citealt{Asano13}; \citealt{Mancini15}).\\
The term regarding dust acccretion $\dot{G}_{i,dust,accr}(t)$ can be expressed in terms of a typical timescale for accretion $\tau_{accr}$:
\begin{equation}
\dot{G}_{i,dust,accr}(t)=\frac{G_{i,dust}(t)}{\tau_{i,accr}}.
\label{e:dust_accr}
\end{equation}
In this way, the problem of finding the dust accretion rate is reduced to find just the typical timescale of accretion. \cite{Hirashita00} expressed the dust accretion timescale for the $i$-th element as follows:
\begin{equation}
\tau_{i,accr}=\frac{\tau_g}{X_{cl} (1 - f_i)}.
\end{equation}
In the latter equation, $f_i=G_{i,dust}/G_i$ (we omit to report the time dependence for convenience) is the DGR for the element $i$ at the time $t$, while $X_{cl}$ represents the mass fraction of molecular clouds in the ISM. $\tau_g$ is the characteristic dust growth timescale. In our models we adopt the relation given by \cite{Asano13}, who expressed the dust growth timescale $\tau_g$ in a molecular cloud as a function of the metallicity $Z$ as:
\begin{equation}
\tau_g=2.0\cdot 10^7yr \cdot \bigg( \frac{Z}{0.02} \bigg)^{-1},
\end{equation}
assuming $50$ $K$ for the cloud temperature, $100$ $cm^{-3}$ for the cloud ambient density and an average value of $0.1$ $\mu m$ for the grain size.

\subsubsection{Dust destruction}
\label{sss:d_destruction}
Dust grains are not only accreted, but experience also destruction in the ISM. The most efficient process among those able to cycle dust
back into the gas phase is the destruction by SN shocks.\\
Similarly to Equation \eqref{e:dust_accr} for dust accretion, $\dot{G}_{i,dust,destr}(t)$ is expressed in terms of the grain destruction timescale $\tau_{destr}$:
\begin{equation}
\dot{G}_{i,dust,destr}(t)=\frac{G_{i,dust}(t)}{\tau_{destr}}.
\label{e:dust_destr}
\end{equation}
This timescale is assumed to be the same for all the elements depleted in dust and has the following form:
\begin{equation}
\tau_{destr}=\frac{M_{ISM}}{(\epsilon \cdot M_{swept}) SN_{rate}},
\end{equation}
where $M_{swept}$ is the ISM mass swept by a SN shock and $\epsilon$ is the efficiency of grain destruction in the ISM. For the last two parameters, in our model the \cite{Asano13} prescriptions are adopted. They suggest an efficiency $\epsilon=0.1$ and predict for the swept mass:
\begin{equation}
M_{swept}=1535 \cdot \big(Z/Z_\odot+0.039\big)^{-0.289} M_\odot,
\end{equation}
assuming $1cm^{-3}$ for the environment. 

\section{Results}
\label{s:results}
In this Section we attempt to identify the main characteristics of the GRB host galaxies of our sample comparing the results given by the chemical evolution models with the abundances measured in the GRB hosts. Following the idea developed in the previous works of \citet{Calura09} and \citet{Grieco14}, the procedure consists in comparing model predictions for $[X/Fe]$ vs. $[Fe/H]$ with the ratios observed in the GRB afterglow spectrum for several chemical elements. In this way it is possible to determine the star formation history and therefore the nature of the hosts.

\subsection{Model specifications}
\label{ss:models}
We ran several models for all the possible morphologies (irregular galaxy, spiral disk, elliptical galaxy) that a GRB host can have, in order to reproduce host galaxy abundance data.
In Table \ref{t:models}, we give a list of the chemical evolution models considered, where the model name is given in the first column. The subsequent columns show the infall mass and timescale, the star formation efficiency (SFE), the wind parameter and the IMF adopted. In the last column, the choice of the condensation efficiencies $\delta_i^{CC}$ for CC-SNe (from \citealt{Piovan11}, see \ref{ss:d_prod}) is specified: $\delta_{HP}$ stands for the condensation efficiencies with a low density circumstellar environment ($n_H = 0.1$ $cm^{-3}$ ), that leads to higher $\delta_i$ values, whereas $\delta_{MP}$ stands for the prescriptions with $n_H = 1cm^{-3}$ density, that leaves lower dust production by massive stars. The reference models for the three different morphological types are written in bold.\\
\begin{figure}
\centering
	\includegraphics[width=\columnwidth]{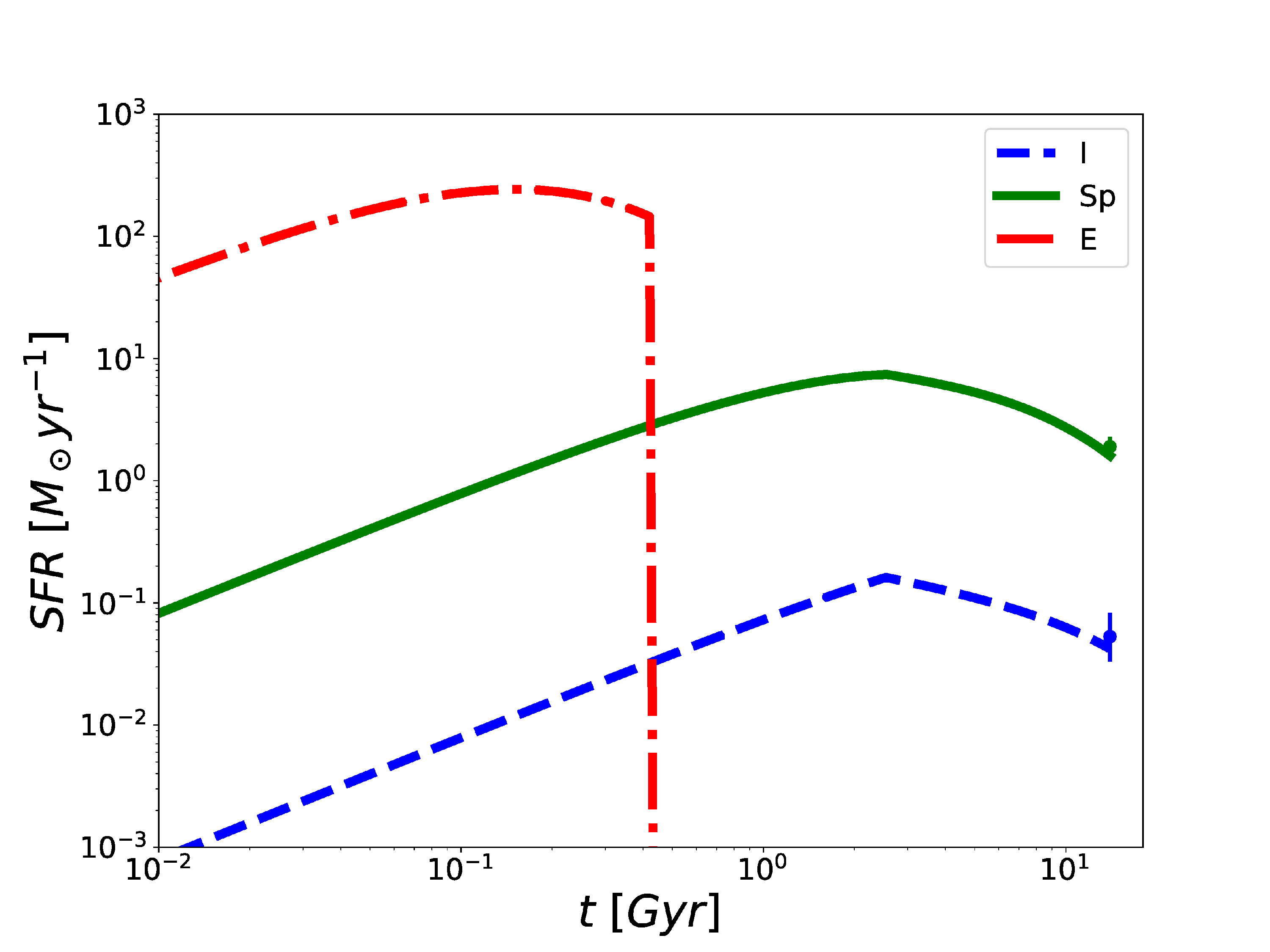}
    \caption{Predicted SFRs behavior with time for the reference models for galaxies of different morphological types. The two points with error bars refer to measured average SFR at the present time in the solar neighborhood (\citealt{Chomiuk11}, in green) and SMC (\citealt{Rubele15}, in blue)}
    \label{f:SFR}
\end{figure}
\begin{figure*}
\centering
\includegraphics[width=.54\textwidth]{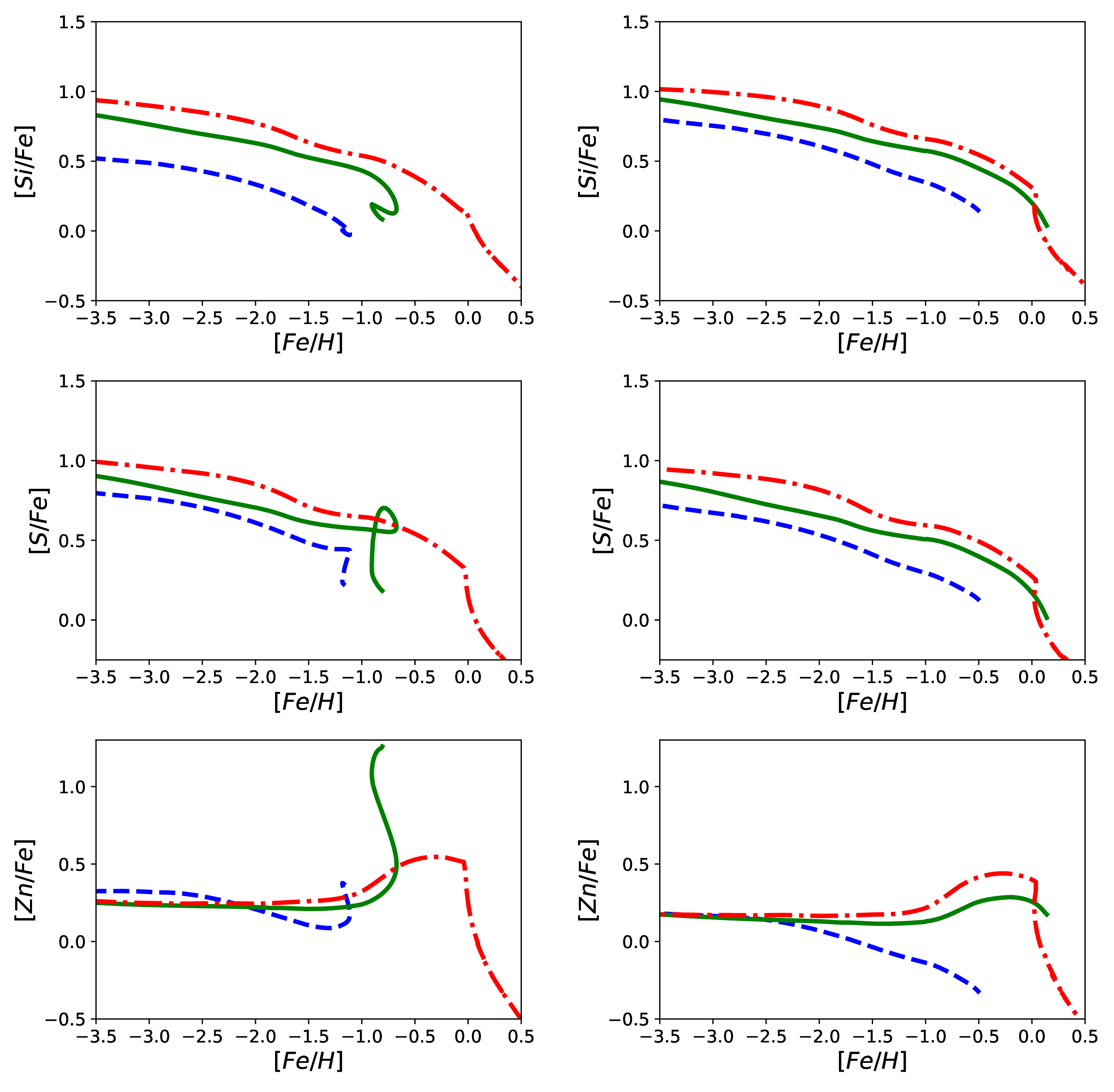}
\centering
\caption{Example of $[X/Fe]$ vs. $[Fe/H]$ ratios behaviour for the chemical evolution models adopted in this work. The blue dashed (I), green solid (Sp) and red dash-dotted lines (E) are the predictions computed by means of reference models for an irregular, a spiral and an elliptical galaxy. The left panels show the models considering dust, whereas the right panels the models without dust.} 
\label{f:GRB_nodust}
\end{figure*}
As we can see from Figure \ref{f:SFR}, the parameters for the reference I and Sp models are fine tuned in order to reproduce the measured average SFR in the Small Magellanic Cloud (SMC, a typical irregular galaxy) and the solar neighbourhood (that represent a spiral disk), respectively. For what concerns the E model, instead, the parameters used trace the typical behaviour of an elliptical galaxy, with a quenching of the star formation, determined by the action of galactic winds, after an initial and very intense burst.
To choose between the different dust condensation efficiencies by massive stars from \citet{Piovan11}, we followed the work of \citet{Gioannini17b}. In this latter paper, the dust prescriptions (which are the same we adopt here) are chosen to reproduce the observed DGRs in local dwarf irregular galaxies and in local spirals, as well as the dust masses observed in high redshift elliptical galaxies (remember \ref{ss:d_prod}). For our reference models we adopt the same prescriptions, since the other chemical evolution model parameters (infall mass, infall timescale, etc.) in the two works are similar.\\
In Table \ref{t:models} we indicate many other models, where the parameters are varied, in order to better identify the host galaxies. These models are codified in a very simple way: every change in the parameters is indicated by a letter or a symbol. In particular, for a modification in the dust production parameters we use a $+$ (higher dust production) or a $-$ (lower dust production), for the adoption of other IMFs with respect to the \citeauthor{Salpeter55} we write the initials of them in capital letter, whereas other changes are signaled with a lowercase letter ($m$ for a mass and SFE increment, $l$ for a mass and SFE decrement).\\
Before starting with the identification, some other model features need to be mentioned. First, during this work we assume for $Ni$ (an element not considered in \citealt{Piovan11}) the same condensation efficiencies as $Fe$. This solution, although approximate, is reasonable, and this is due to the very similar condensation temperatures (\citealt{Taylor01}) and the fact that $Ni$ belongs to the so-called $Fe$-peak group of elements. Thanks to this, in our work we consider $C$, $O$, $Mg$, $Si$, $S$, $Fe$ and $Ni$ as refractory elements (i.e. apt to be condensed in dust). On the other hand, for $Zn$ and $N$, we assume no dust depletion, since it is known that the two elements are volatile, with very little variation (up to $0.1$ $dex$) between gas and total abundances. In Figure \ref{f:GRB_nodust} we show what happens in considering or not dust in our models.
Regarding $N$, in this paper we ran all the models twice. First we adopted \citet{Nomoto13} yields for $N$, which do not consider primary production from massive stars, and then we use \citet{Matteucci86} prescription, considering instead a fixed amount of primary $N$ produced by this kind of stars. In this way we can also possibly better understand the $N$ primary production issue, exploiting the $N$ abundance data from GRB hosts.

\subsection{Host identification}
\label{ss:host_id}
In order to constrain the nature of the host galaxies analysed, we used chemical evolution models able to account for the different behaviour of $[X/Fe]$ vs. $[Fe/H]$ patterns in different galaxy types (see Section \ref{s:intro}).\\
To determine which of the galaxy models is the best in reproducing the abundance data, we adopted a statistical test, already used in the works of \citeauthor{Dessauges04} (\citeyear{Dessauges04}, \citeyear{Dessauges07}). This method is particularly important when the best solution cannot be clearly identified at first sight. This test consists in determining the minimal distance between the data point and the curve of the model for each abundance diagram $[X/Fe]$ vs. $[Fe/H]$ we have for a host. In particular, we derived this minimal distance by looking for the distance $d_X$ for which the ratio $d_X/\sigma_X$, where $\sigma_X$ is the error for the abundance data, is minimal. After that, we computed the weighted mean for all the abundance diagrams considered in each system. From the comparison of these means, we obtained the best model in representing the GRB host. This procedure also gives the opportunity to approximately estimate the age of the host galaxy. Each point of minimal distance inferred for the best model has in fact a time $t_X$. By weighting these times for the reciprocals of the ratios $d_X/\sigma_X$, we derived the age of the host. We signal that upper and lower limits are not taken into account in this procedure.
\begin{figure*}
\centering
\includegraphics[width=.81\textwidth]{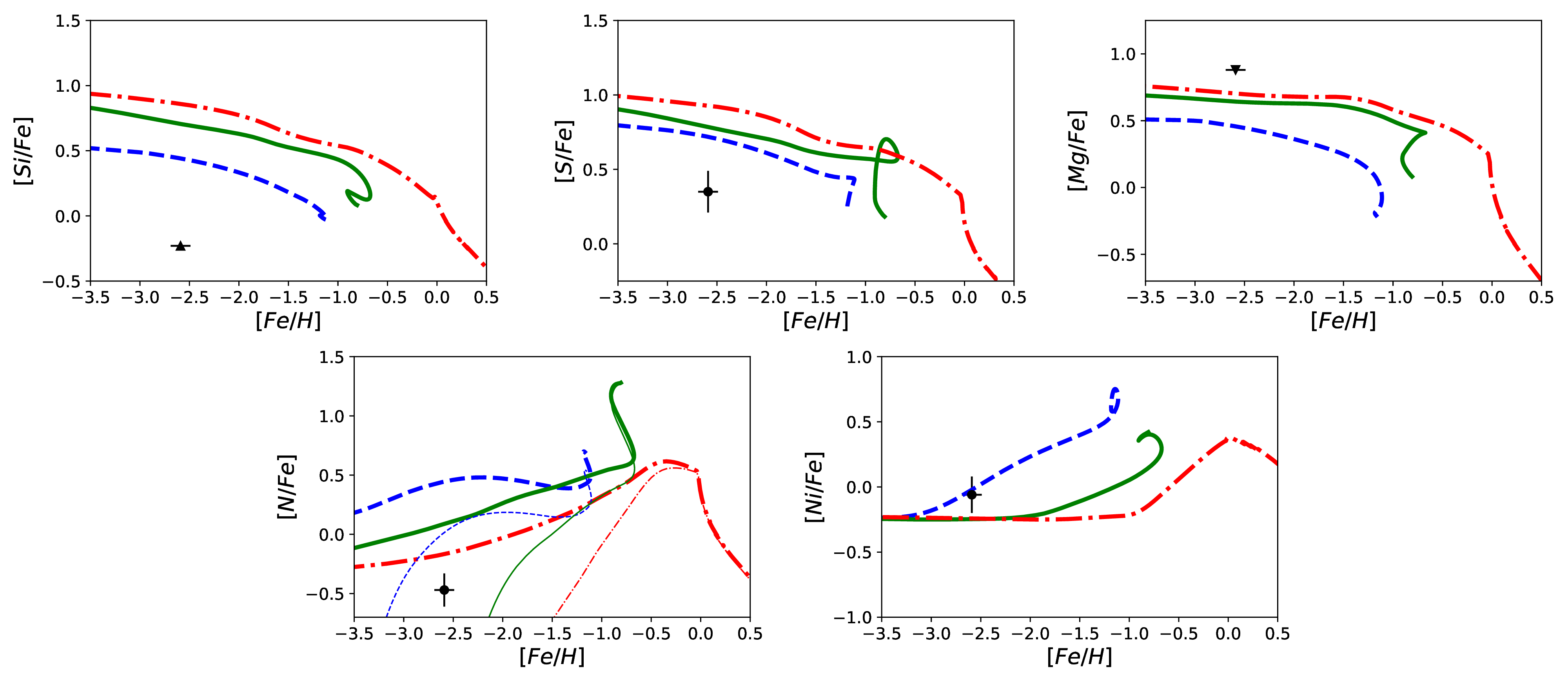}
\centering
\caption{Observed $[X/Fe]$ vs. $[Fe/H]$ ratios for the GRB 050730 host galaxy provided by \citet{Prochaska07}. Up and down arrows indicate lower and upper limits data. The blue dashed (I), green solid (Sp) and red dash-dotted lines (E) are the predictions computed by means of reference models for an irregular, a spiral and an elliptical galaxy. In the lower left panel are shown the results considering primary production from massive stars (\citealt{Matteucci86}, thick lines) and considering \citet{Nomoto13} (thin lines) yields for $N$, respectively.} 
\label{f:050730}
\end{figure*}
\begin{figure*}
\centering
\includegraphics[width=.81\textwidth]{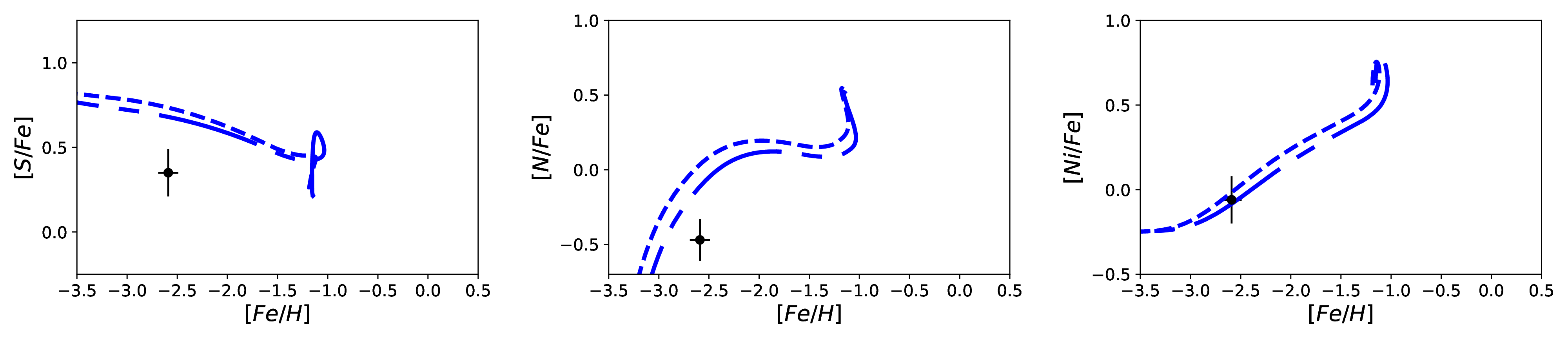}
\caption{Observed $[X/Fe]$ vs. $[Fe/H]$ ratios for the GRB 050730 host galaxy provided by \citet{Prochaska07}. The blue dashed (I) and blue long-short dashed line (I$-$) are the prediction computed by means of the reference model for an irregular galaxy and the model for an irregular with decremented dust production from CC-SNe, respectively. In the central panel are shown the results considering \citet{Nomoto13} yields for $N$.} 
\label{f:050730_1}
\end{figure*}
Before starting, we have to say that model results for $Ni$ and $Zn$ should be taken with caution, especially because their stellar yields are still quite uncertain. As a matter of fact, we can look at the results by \citet{Koba06} of chemical evolution models for $Ni$ in the solar neighbourhood, adopting \citet{Koba06} yields for massive stars (which are very similar to those of \citealt{Nomoto13}). At the same time $Zn$ yields are still affected by quite large uncertainties, with many hypotheses formulated and discarded in the past years on its production (e.g. type Ia SNe by \citealt{Matteucci93}).

\subsubsection{GRB 050730}
The data-models comparison at Figure \ref{f:050730} show quite good agreement with the irregular galaxy reference model. In particular, $[S/Fe]$ and the $[Ni/Fe]$ data are the main drivers of this hypothesis, corroborated by the compatibility with $Si$ and $Mg$ lower and upper limits. Concerning $[N/Fe]$, the observed ratio is more in agreement with the proposed explanation in the case we consider only secondary production by massive stars (\citealt{Nomoto13}, thin lines). Considering instead primary production (\citealt{Matteucci86}, thick lines) the models passes too high to agree with the observations. For this reason in our statistical test we consider only models with secondary $N$ production.\\
From this test, we determined the irregular galaxy model with decreased dust production by massive stars (I$-$) as the best model among those shown in Table \ref{t:models}. In Figure \ref{f:050730_1} are shown the patterns of this model, together with the reference model for irregular galaxies. The better agreement is evident watching in particular the panels for $[N/Fe]$ and $[S/Fe]$. For what concerns this latter ratio, we have to say that the observed ratio is even better reproduced by the irregular low mass ($< 10^9M_\odot$) and SFE ($<0.05$ $ Gyr^{-1}$) model (I$l$). However, at the same time we have worse agreement looking at $[Ni/Fe]$ and mostly at $[N/Fe]$, even if we have to remind the uncertainties in the production process of $N$, that can possibly alter the results.\\
In this way, adopting the less dusty \textbf{irregular} model as the best one, we estimated the age of the host at the time of the GRB event. We found for it an age of $\sim 0.2$ $Gyr$.

\subsubsection{GRB 050820}
\begin{figure*}
\centering
\includegraphics[width=.81\textwidth]{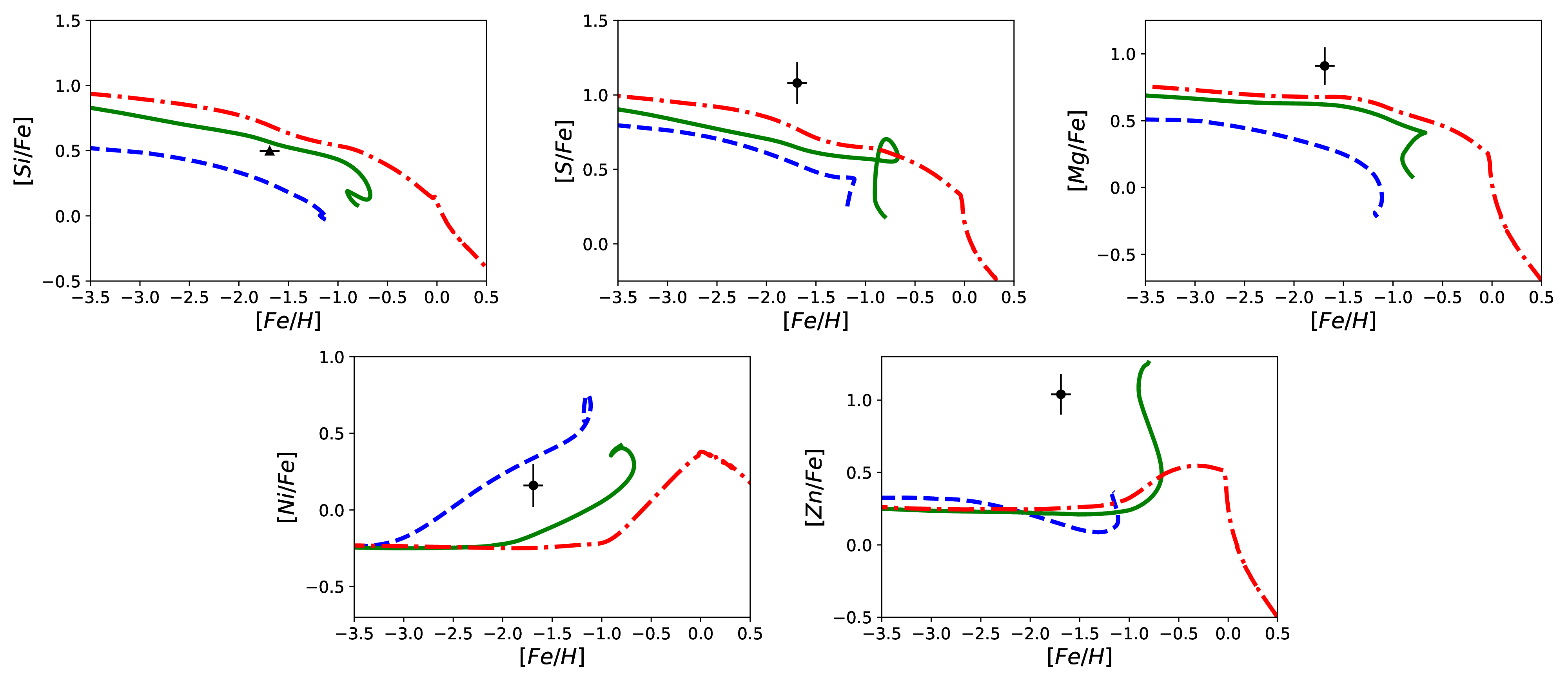}
\centering
\caption{Observed $[X/Fe]$ vs. $[Fe/H]$ ratios for the GRB 050820 host galaxy provided by \citet{Prochaska07}. Up arrows indicate lower limit data. The blue dashed (I), green solid (Sp) and red dash-dotted line (E) are the predictions computed by means of reference models for an irregular, a spiral and an elliptical galaxy, respectively.} 
\label{f:050820}
\end{figure*}
\begin{figure*}
\centering
\includegraphics[width=.81\textwidth]{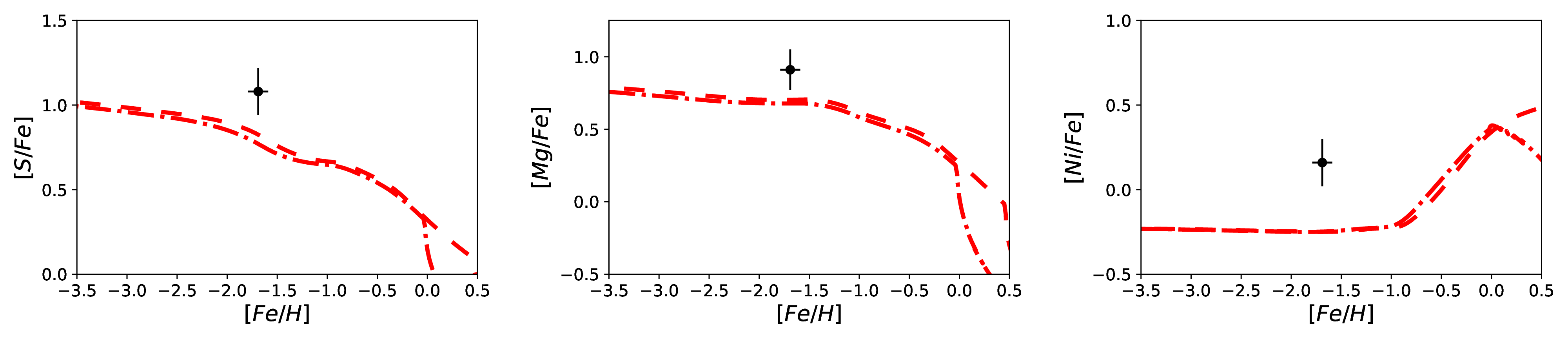}
\centering
\caption{Observed $[X/Fe]$ vs. $[Fe/H]$ ratios for the GRB 050820 host galaxy provided by \citet{Prochaska07}. The red dash-dotted (E) and red dashed line (E$m$) are the predictions computed by means of reference models for an elliptical galaxy and an elliptical model with increased mass and SFE, respectively.} 
\label{f:050820_1}
\end{figure*}
\begin{figure*}
\centering
\includegraphics[width=.81\textwidth]{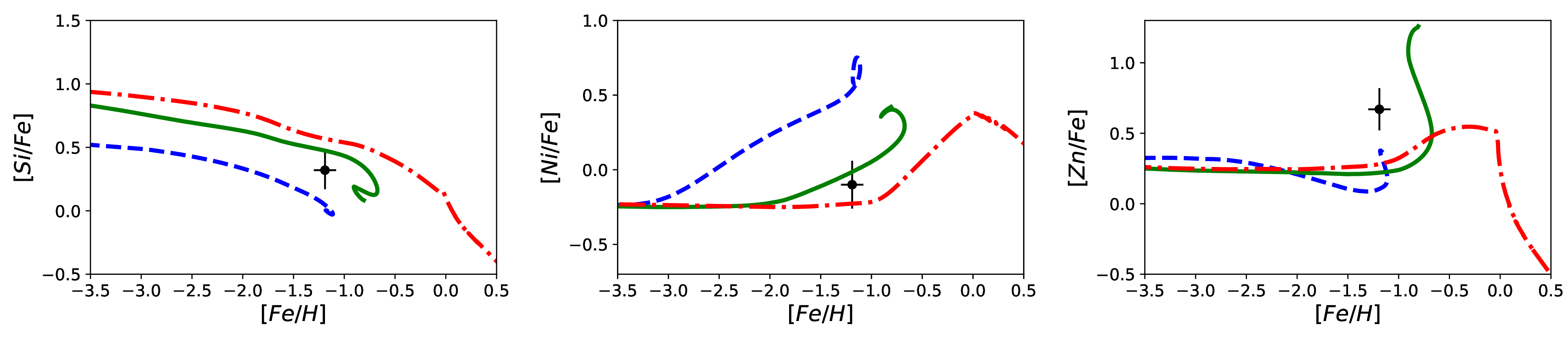}
\centering
\caption{Observed $[X/Fe]$ vs. $[Fe/H]$ ratios for the GRB 081008 host galaxy provided by \citet{DElia11}. The blue dashed (I), green solid (Sp) and red dash-dotted line (E) are the predictions computed by means of reference models for an irregular, a spiral and an elliptical galaxy, respectively.} 
\label{f:081008}
\end{figure*}
\begin{figure*}
\centering
\includegraphics[width=.81\textwidth]{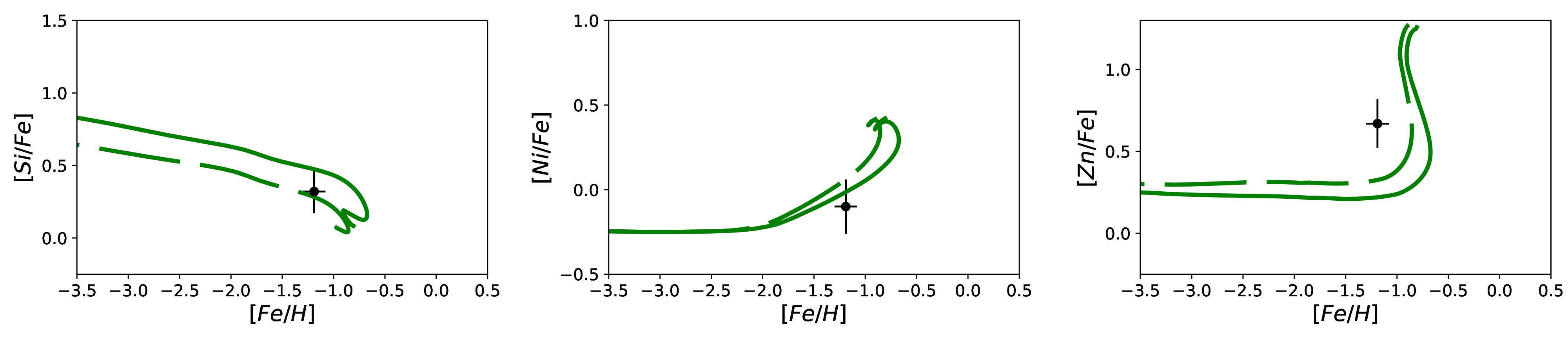}
\caption{Observed $[X/Fe]$ vs. $[Fe/H]$ ratios for the GRB 081008 host galaxy provided by \citet{DElia11}. In the first graph,the green solid (Sp) and green long dashed line (Sp$+$) are the predictions computed by means of the reference model for a spiral disk and the model for a spiral with incremented dust production from CC-SNe, respectively.} 
\label{f:081008_1}
\end{figure*}
\begin{figure*}
\centering
\includegraphics[width=.81\textwidth]{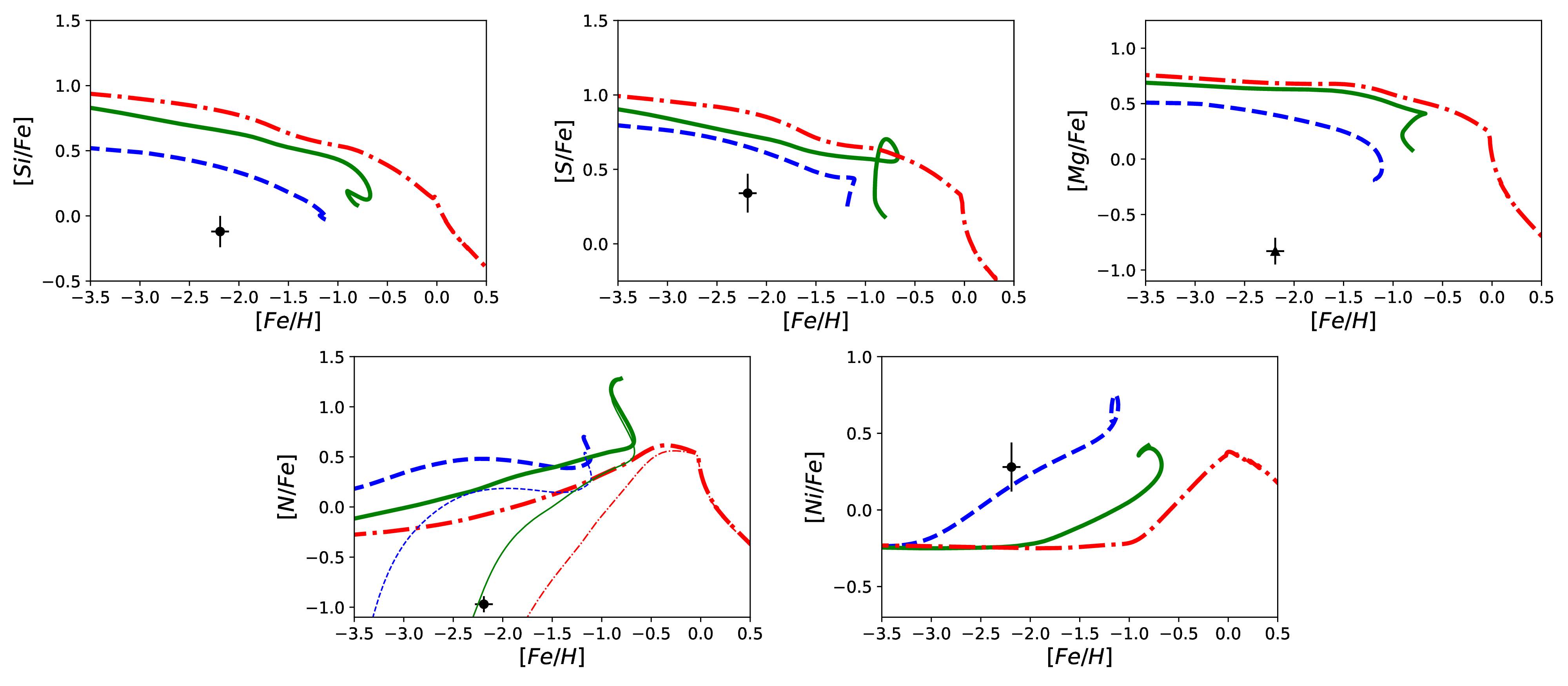}
\centering
\caption{Observed $[X/Fe]$ vs. $[Fe/H]$ ratios for the GRB 090926A host galaxy provided by \citet{DElia10}. Up arrows indicate lower limit data. The blue dashed (I), green solid (Sp) and red dash-dotted line (E) are the predictions computed by means of reference models for an irregular, a spiral and an elliptical galaxy. In the lower left panel are shown the results considering primary production from massive stars (\citealt{Matteucci86}, thick lines) and considering \citet{Nomoto13} (thin lines) yields for $N$, respectively.} 
\label{f:090926}
\end{figure*}
\begin{figure*}
\centering
\includegraphics[width=.81\textwidth]{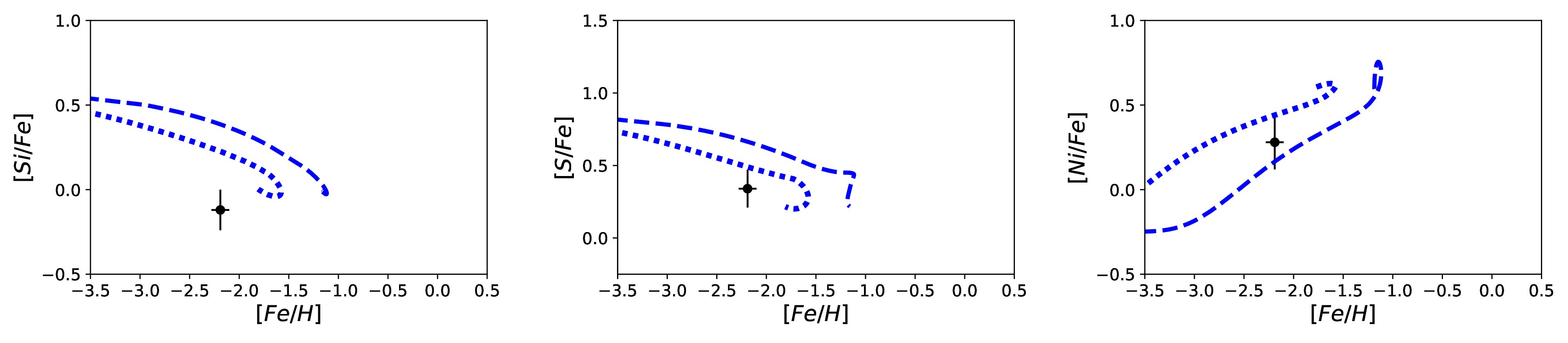}
\caption{Observed $[X/Fe]$ vs. $[Fe/H]$ ratios for the GRB 090926A host galaxy provided by \citet{DElia10}. The bue dashed (I) and blue dotted line (I$l$) are the prediction computed by means of the reference model for an irregular galaxy and the model for an irregular with lower infall mass and SFE, respectively} 
\label{f:090926_1}
\end{figure*}
From Figure \ref{f:050820}, we see that the three upper panels for $[\alpha/Fe]$ ratios indicate for this host an elliptical galaxy. At the same time, the panel showing $[Ni/Fe]$ vs. $[Fe/H]$ behaviour has better agreement for late-type (irregular, spiral) galaxy models. However, we remind the uncertainties (see \ref{ss:host_id}) we have in $Ni$ yields, in order to explain this discrepancy. For $[Zn/Fe]$, instead, we note that the observed value is much higher than all the model predictions. We will see however that the overabundance of this ratio with respect to what is predicted by the models is a common feature for all the host studied for which we have data. Rather, we can exploit the very high $[Zn/Fe]$ value ($>1$ $dex$) in this host as a further insight for the elliptical galaxy hypothesis. We observe indeed a correlation between the $Zn$ and the $\alpha$-element abundances, where $[Zn/Fe]$ ratios are higher when $[\alpha/Fe]$ ratios are also. In other words, $Zn$ seems to behave like $\alpha$-elements.\\
In Figure \ref{f:050820_1} we show what happens adopting the model E$m$ for an elliptical galaxy with increased mass ($10^{12}M_\odot$) and SFE ($25$ $Gyr^{-1}$), which results the best one from the statistical test performed. As expected from the $\lq\lq$time-delay model", $[\alpha/Fe]$ ratios tend to rise in increasing mass and SF. This of course help the E$m$ model to be more consistent with the observational data. For $Ni$, instead, the situation is quite the same. Note that we do not show $[Zn/Fe]$ vs. $[Fe/H]$ plot in this Figure, due to the large discrepancy between the observed $[Zn/Fe]$ and all the model tracks.\\
Concluding, we identify this host galaxy as a massive, strong star forming \textbf{elliptical} galaxy. For what concerns the age of this host, we found for it a very young one of $\sim 15$ $Myr$.

\subsubsection{GRB 081008}
For this host, as shown in Figure \ref{f:081008}, the observed abundance ratios are quite well fitted by the reference models for spiral galaxies, in particular looking at $[Si/Fe]$ and $[Ni/Fe]$ panels. Also the not so high $[Zn/Fe]$ ratio ($\sim 0.6$ $dex$) corroborate the idea of having a late-type galaxy (as the spiral is). We remind in fact the $\alpha$-$Zn$ correlation claimed in the analysis of GRB 050820 host galaxy.\\
Performing our test, the resulting best model was found to be the one of for spiral galaxies with increased dust production by massive stars (Sp$+$ model). In Figure \ref{f:081008_1} are shown the results for this model, together with those of the reference spiral model. As we can see, we have a big improvement in particular in fitting the $[Si/Fe]$ ratio data. We note also a better fit of the $[Zn/Fe]$ observed ratio by Sp$+$ model, despite of not passing in the error bars. However, some consideration can be drawn in this case. In fact, the relatively high gas metallicity observed ($[Fe/H]=-1.19$ $dex$), could explain the $[Zn/Fe]$ value also in terms of dust accretion. Reducing by a small factor the typical accretion timescale adopted in this work (from \citealt{Asano13}), we can obtain a model track passing for the observed ratio. For what concerns $Ni$, we have a bit worse agreement for the Sp$+$ model. However, this is small compared to the improvement we found for the other abundace ratios.\\
Adopting the dusty \textbf{spiral} model as the best one for this host galaxy, we estimate the age of the host at the time of the GRB event to be $\sim 0.45$ $Gyr$.

\subsubsection{GRB 090926A}
\begin{figure*}
\centering
\includegraphics[width=.81\textwidth]{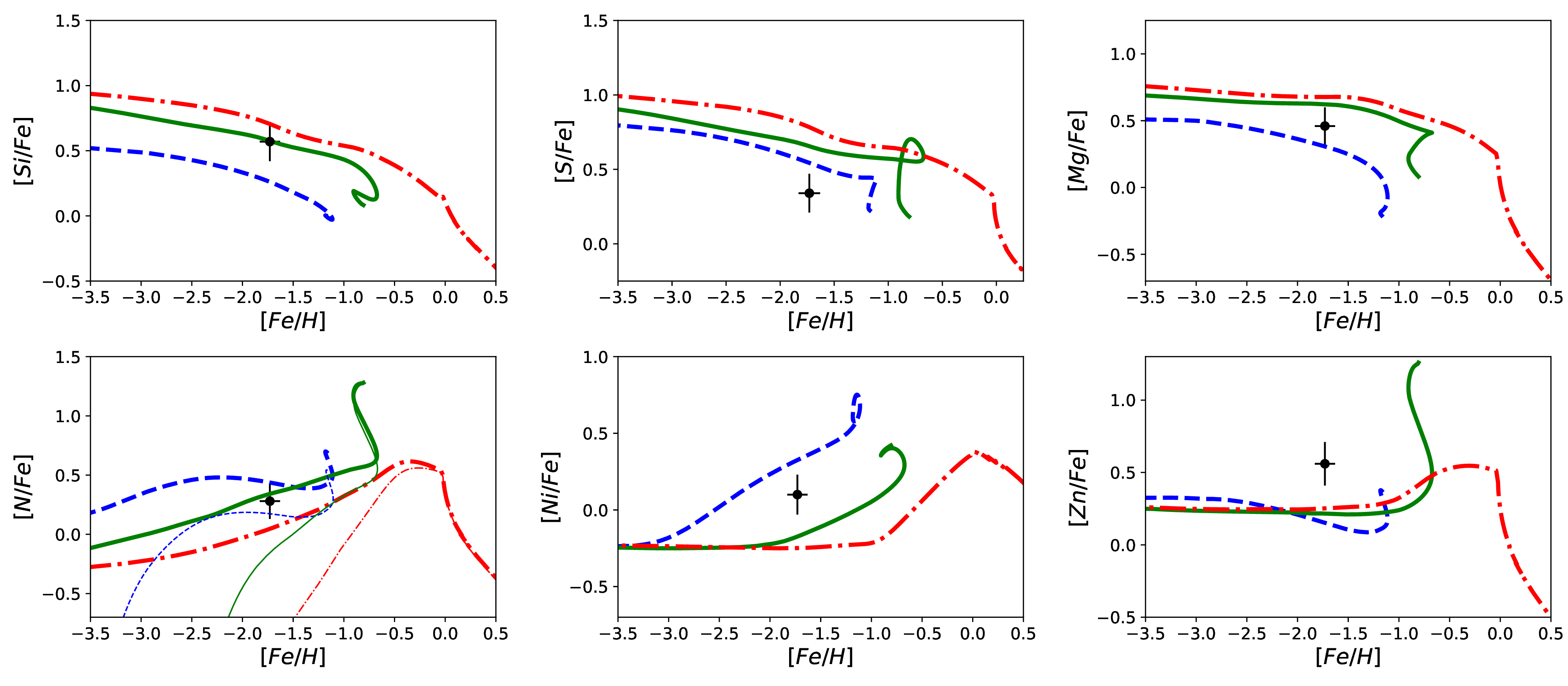}
\centering
\caption{Observed $[X/Fe]$ vs. $[Fe/H]$ ratios for the GRB 120327A host galaxy provided by \citet{DElia14}. The blue dashed (I), green solid (Sp) and red dash-dotted line (E) are the predictions computed by means of reference models for an irregular, a spiral and an elliptical galaxy. In the lower left panel are shown the results considering primary production from massive stars (\citealt{Matteucci86}, thick lines) and considering \citet{Nomoto13} (thin lines) yields for $N$, respectively.} 
\label{f:120327}
\end{figure*}
\begin{figure*}
\centering
\includegraphics[width=.81\textwidth]{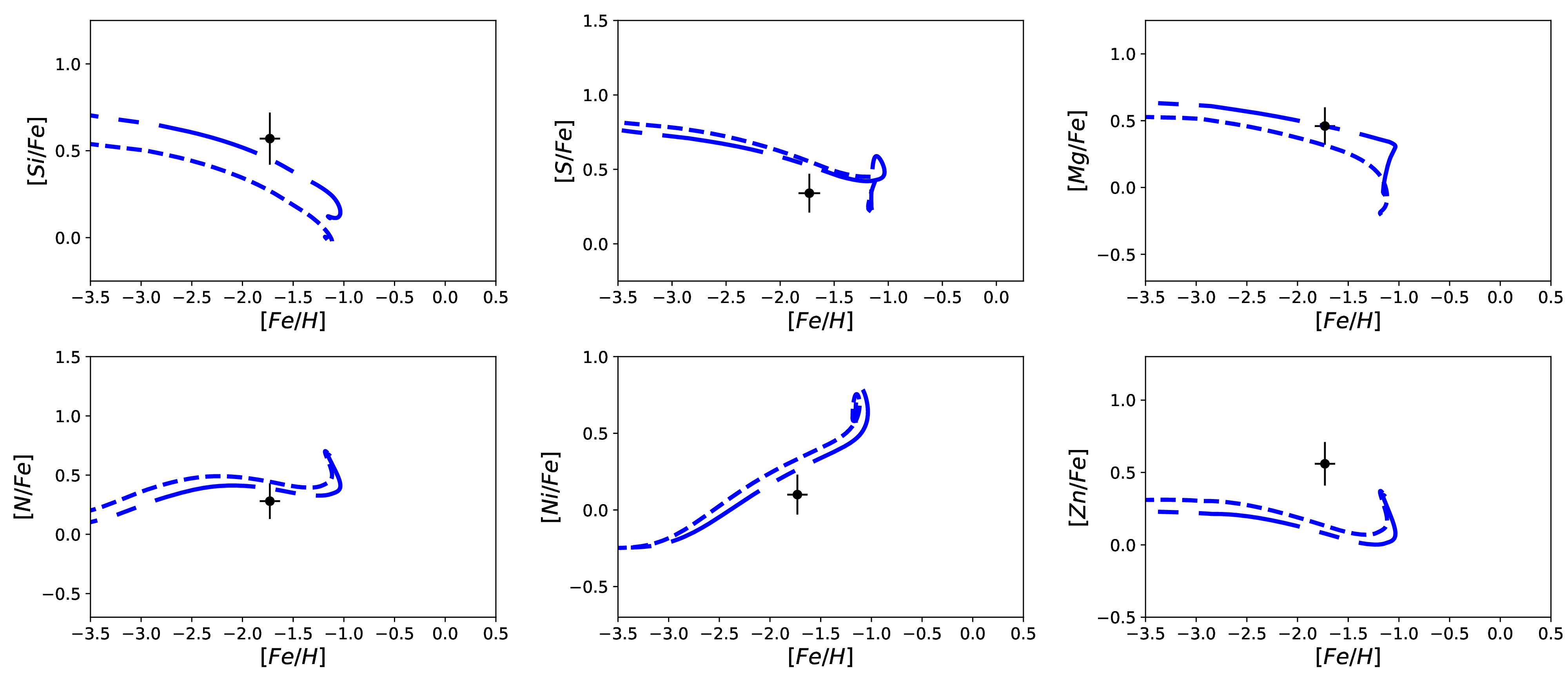}
\caption{Observed $[X/Fe]$ vs. $[Fe/H]$  ratios for the GRB 120327A host galaxy provided by \citet{DElia14}. The blue dashed (I) and blue long-short dashed line (I$−$) are the predictions computed by means of the reference model for an irregular galaxy and the model for an irregular with decremented dust production from CC-SNe. In the lower left panel are shown the results considering primary production from massive stars (\citealt{Matteucci86}) for $N$.}
\label{f:120327_1}
\end{figure*}
Looking at Figure \ref{f:090926}, the three upper panels showing $[\alpha/Fe]$ vs. $[Fe/H]$ behaviours show very low abuandance values for $\alpha$-elements (however, $Mg$ is a lower limit). This suggests ($\lq\lq$time-delay model") we have to deal with an irregular galaxy. Also $[Ni/Fe]$ plot in the lower right panel corroborate such hypothesis, since it is in good agreement with the reference irregular model. The lower left panel of Figure \ref{f:090926} shows instead the predicted $[N/Fe]$ behaviour in cases of primary (\citealt{Matteucci86}) or secondary (\citealt{Nomoto13}) production by massive stars. In this case the abundance data agree with the spiral model considering \citet{Nomoto13} yields, whereas it does not at all considering primary production. However, the paper from which we derived the abundance data (\citealt{DElia10}) highlights problems in its determination. For this reason, we excluded this element from our statistical analysis.\\
Among all the models considered in Table \ref{t:models}, we found that the low mass and SF irregular one (I$l$) is the best to describe this host. In Figure \ref{f:090926_1} are shown the results for this latter and the irregular reference model. It is evident that we have better agreement with data lowering mass and star formation, in particular looking at $[Si/Fe]$ and $[S/Fe]$ in left and central panel. Still looking at Figure \ref{f:090926_1}, we see that $[S/Fe]$ observed ratio is more in agreement with the I$l$ model than $[Si/Fe]$. Even better, we note in general that models for $S$ are lower with respect to the data comparing to what happens to $Si$. This can be explained with an underestimation of the amount of dust in the galaxy. In fact, due to the different condensation efficiencies for $Si$ (higher) and $S$ (lower), we expect that in a more dusty galaxy $[S/Fe]$ tends to be higher, whereas $[Si/Fe]$ lower, due to the different dust depletion of these elements with respect to $Fe$. For the irregular models adopted here we set the highest possible condensation efficiencies for CC-SNe $\delta_{HP}$, to follow the results of \citet{Gioannini17b} on local irregular galaxies, so we cannot test properly what just said. However, looking at what happens in spiral galaxies increasing the dust production rate by CC-SNe, the hypothesis of having a very dust rich host can be considered reliable.\\
Since it resulted the best model, we adopt the low mass ($<10^9M_\odot$) and SFE ($<0.05$ $Gyr^{-1}$) \textbf{irregular} galaxy model to estimate the galaxy age. We found for this host galaxy an age $\sim 1.15$ $Gyr$. This makes the host of GRB 090926A the oldest in terms of galactic age among those studied in this work.

\subsubsection{GRB 120327A}
\begin{figure*}
\centering
\includegraphics[width=.81\textwidth]{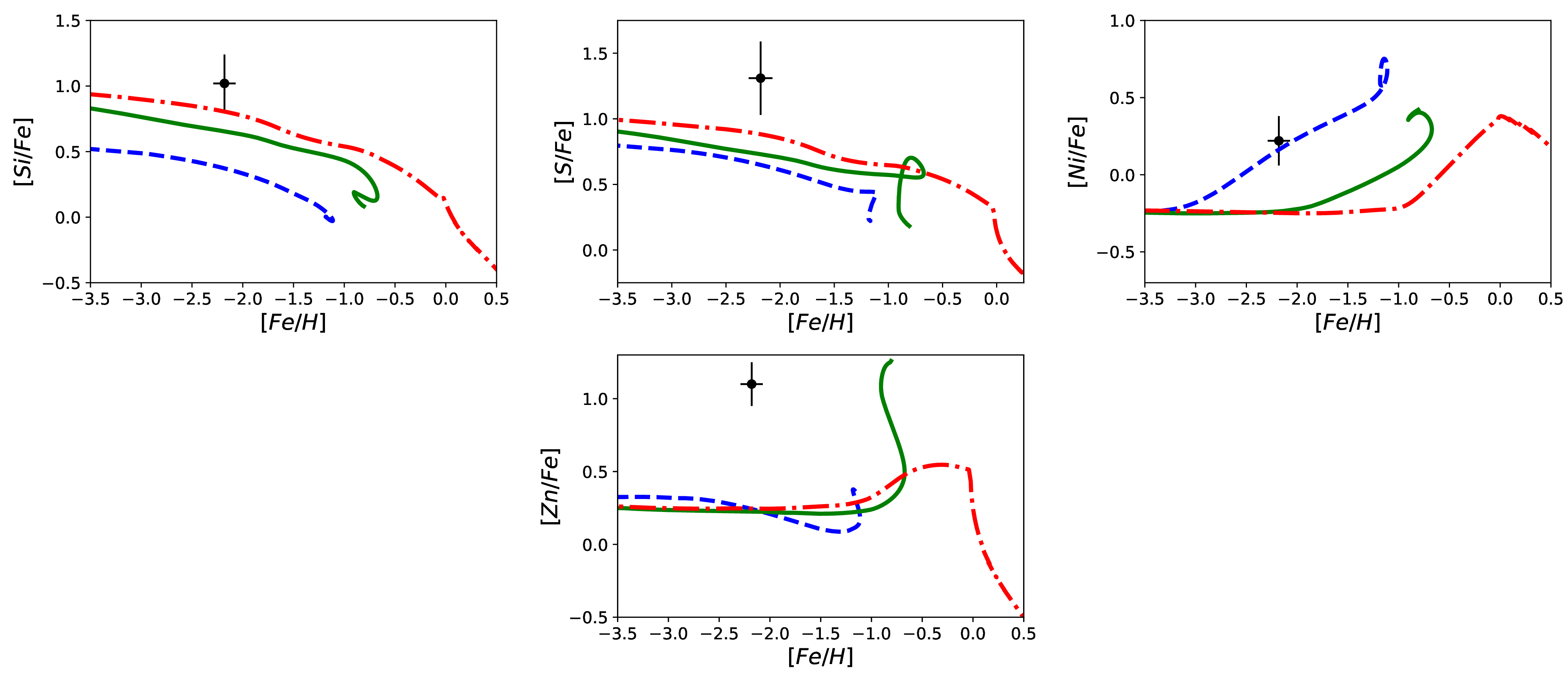}
\centering
\caption{Observed $[X/Fe]$ vs. $[Fe/H]$ ratios for the GRB 120815A host galaxy provided by \citet{Kruhler13}. Up and down arrows indicate lower and upper limits data. The blue dashed (I), green solid (Sp) and red dash-dotted line (E) are the predictions computed by means of reference models for an irregular, a spiral and an elliptical galaxy, respectively.} 
\label{f:120815}
\end{figure*}
\begin{figure*}
\centering
\includegraphics[width=.81\textwidth]{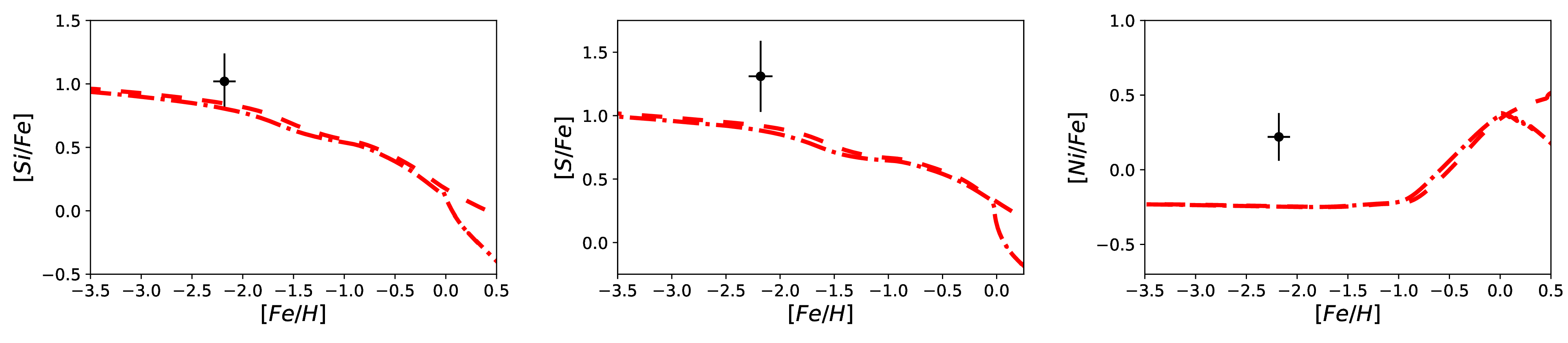}
\caption{Observed $[X/Fe]$ vs. $[Fe/H]$ ratios for the GRB 120815A host galaxy provided by \citet{Kruhler13}. Up and down arrows indicate lower and upper limits data. The red dash-dotted (E) and red dashed line (E$m$) are the predictions computed by means of the reference model for an elliptical galaxy and an elliptical model with increased mass and SFE, respectively.}
\label{f:120815_1}
\end{figure*}
The comparison of the patterns for the three reference models with the observed abundances in Figure \ref{f:120327} shows quite different results depending on the elements. Looking at the $\alpha$-elements we have that $[Si/Fe]$ is perfectly coherent with the spiral disk model. On the other hand, $[Mg/Fe]$ stays between spiral and irregular models, whereas $[S/Fe]$ indicates an irregular galaxy. In the lower left panel are plotted both the cases in which we consider primary production by massive stars (\citealt{Matteucci86}) and only secondary production (\citealt{Nomoto13}) for $N$. In the first case abundance data are compatible with both the irregular and the spiral reference models. In the case of secondary production, instead, only the irregular model becomes acceptable. $[Ni/Fe]$ observed ratio stays between the patterns for irregular and spiral galaxy models. For what concerns $Zn$, the $[Zn/Fe]$ ratio is higher to what predicted by our models (we remind the uncertainties in $Zn$ yields). However, we have a relatively low abundance with respect to the other $[Zn/Fe]$ data considered in the study. This low abundance ratio strengthen both our first guess of a late type host and that there has to be the previously claimed $\alpha$-$Zn$ correlation.\\
The issue of the different behaviour of the various elements is well explained by the different behaviour of these with dust. In particular, we note a higher abundance value (than predicted by the models) for the element more apt to be condensed in dust ($Si$) and progressively lower values for these ones which are less condensed ($Mg$ and $S$). This scenario is compatible with an environment with less dust content than what we have in the reference models. By means of our test, in fact, we found that the irregular model with decreased dust production by CC-SNe (I$-$) is the best in explaining this GRB host. In Figure \ref{f:120327_1}, we show this latter model together with the irregular reference one. It can be noted better agreeement not only for $\alpha$-elements, but also for $N$ and $Ni$. $[Zn/Fe]$ plot does not show improvements, but we remind the uncertainties we have in the yield computation.\\
In conclusion, we identified this host galaxy as a not so dusty (with respect to the average of this morphological type) \textbf{irregular} galaxy. In this way, we found that the age for this host at the time of the GRB event is $\sim 0.8$ $Gyr$.

\subsubsection{GRB 120815A}
\begin{figure*}
\centering
\includegraphics[width=.81\textwidth]{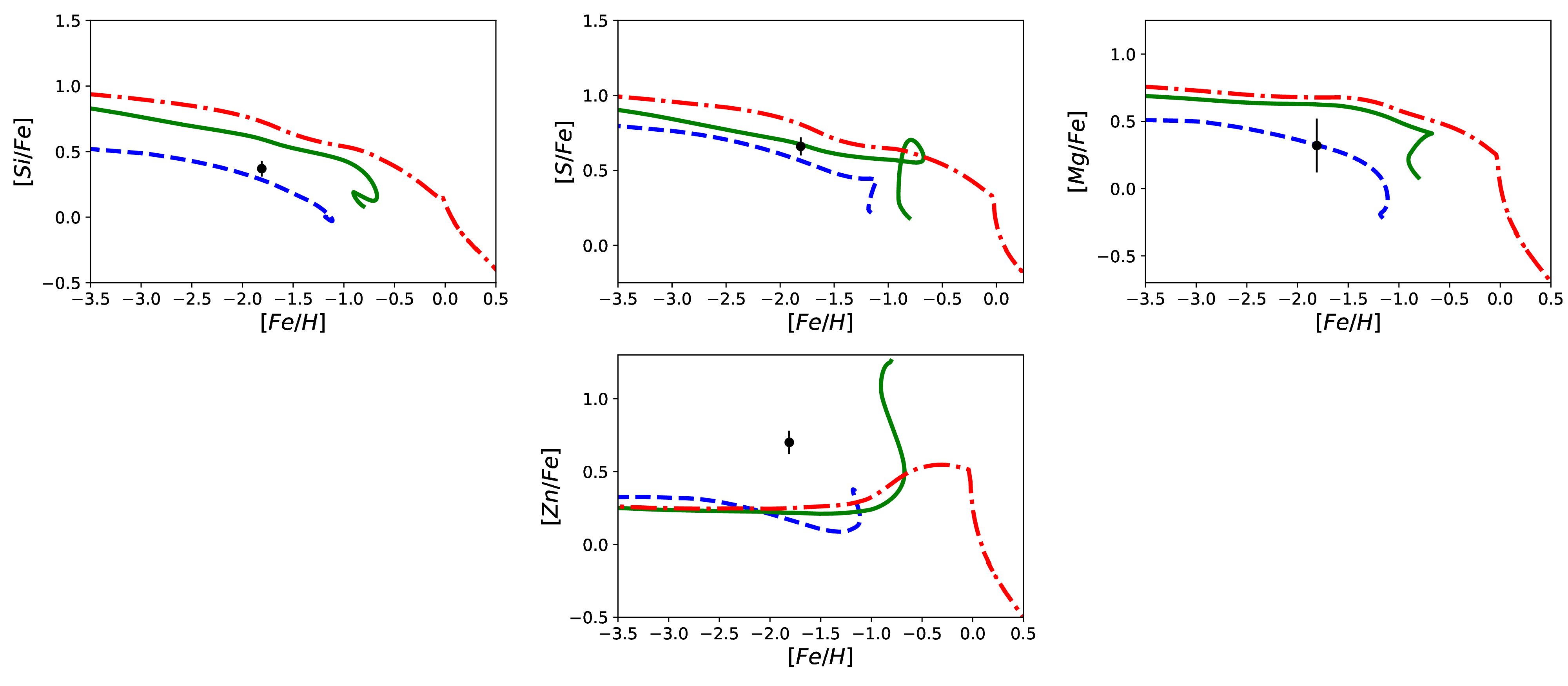}
\centering
\caption{Observed $[X/Fe]$ vs. $[Fe/H]$ ratios for the GRB 161023A host galaxy provided by \citet{deUgarte18}. The blue dashed (I), green solid (Sp) and red dash-dotted line (E) are the predictions computed by means of reference models for an irregular, a spiral and an elliptical galaxy, respectively.} 
\label{f:161023}
\end{figure*}
\begin{figure*}
\centering
\includegraphics[width=.81\textwidth]{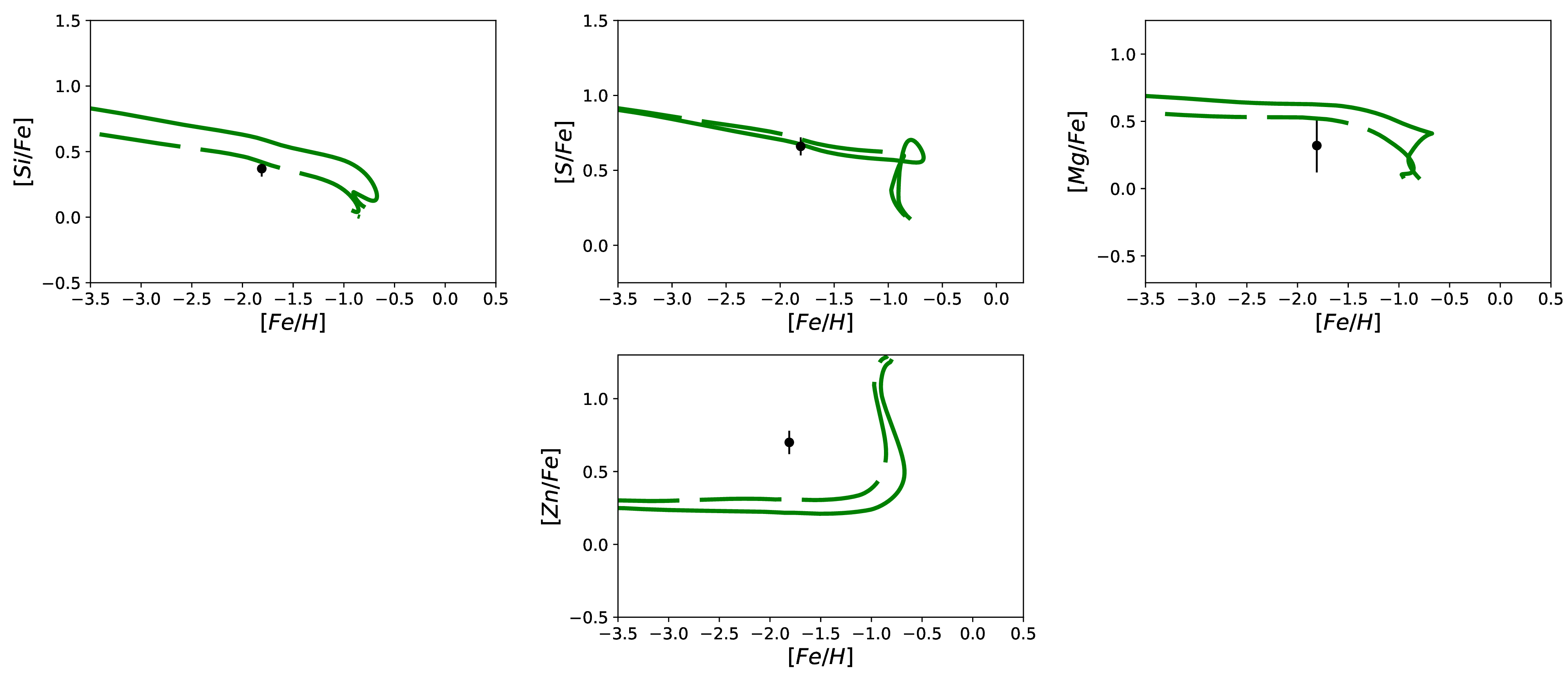}
\caption{Observed $[X/Fe]$ vs. $[Fe/H]$ ratios for the GRB 161023A host galaxy provided by \citet{deUgarte18}. The green solid (Sp) and green long dashed line (Sp$+$) are the predictions computed by means of the reference model for a spiral disk and the model for a spiral with incremented dust production from CC-SNe, respectively.}
\label{f:161023_1}
\end{figure*}
In Figure \ref{f:120815} are compared the abundance patterns for the reference models with the observed data for this host. The $[\alpha/Fe]$ ratios (where $\alpha$ are $Si$ and $S$) are very high: this feature is an indicator of an early type galaxy. Such hypothesis tends to be confirmed by the $[Zn/Fe]$ observed ratio, due to the claimed $\alpha$-$Zn$ correlation (see GRB 050820). However, for the unagreement with the data of all the $[Zn/Fe]$ patterns, we did not considered such element in our statistical analysis. Concerning the observed $[Ni/Fe]$, instead, the plot shows good agreement with the irregular galaxy model. We remind however the uncertainties in $Ni$ models to explain the different behaviour.\\
As shown in Figure \ref{f:120815_1}, increasing the mass and the SFE was found to be the right direction to move, since it increases $[\alpha/Fe]$. The massive ($10^{12}M_\odot$) and strongly star forming ($\nu=25$ $Gyr^{-1}$) \textbf{elliptical} galaxy model (E$m$), in fact, was resulted to be the best in explaining this GRB host galaxy, though the observed $[S/Fe]$ remains still too high than what predicted. In this scenario, the host age was found to be $\sim 10$ $Myr$.

\subsubsection{GRB 161023A}
In Figure \ref{f:161023}, the reference model-data comparison highlights different behaviours for what concerns the various $[\alpha/Fe]$ ratios. $[S/Fe]$ data is in good agreement with the spiral disk model, whereas $Si$ and $Mg$ tend to suggest an irregular galaxy. As for the other hosts studied, $[Zn/Fe]$ stays above the models of the three galaxy types. However, the observed abundance is compatible with other $[Zn/Fe]$ values observed in identified late-type hosts.\\
The behaviour found for different $\alpha$-elements is quite well explained by the dusty (i.e. increased dust production by CC-SNe) spiral model (Sp$+$), shown in Figure \ref{f:161023_1}. Despite of it does not perfectly agree with $[Mg/Fe]$ observed ratio, the model was resulted the best to explain at the same time $Si$ and $S$, characterised by different behaviours with dust.\\
Indeed, the statistical test confirm our explantion, choosing the dusty \textbf{spiral} disk scenario as the favourite to explain this host. Concerning the host age, it was estimated to be $\sim 0.15$ $Gyr$ at the time of the GRB event.
\begin{figure*}
\centering
\includegraphics[width=.54\textwidth]{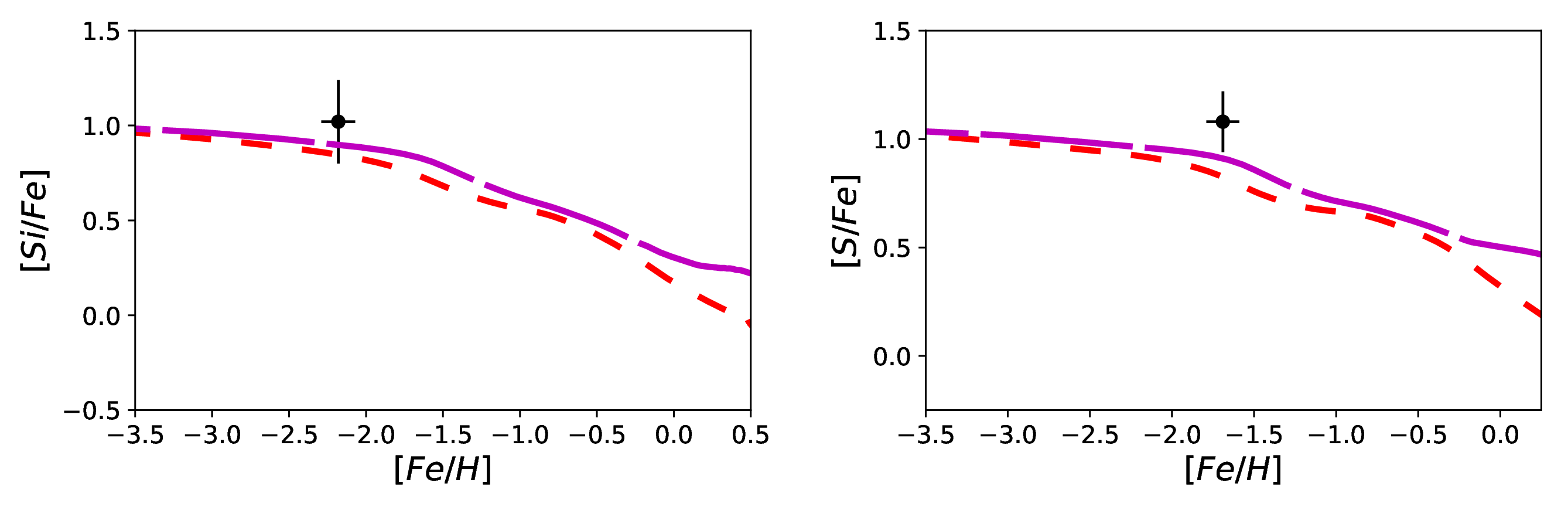}
\caption{Observed $[X/Fe]$ vs. $[Fe/H]$ ratios for the GRB 120815A (left panel) and GRB 050820 (right panel) host galaxies. The red dashed (E$m$) and magenta very long dashed (ET$m$) are the predictions computed by means of the model for an elliptical galaxy with increased mass and SFE and an elliptical model with increased mass, SFE and a top-heavy IMF, respectively.}
\label{f:top_heavy}
\end{figure*}
\begin{figure*}
\centering
\includegraphics[width=.54\textwidth]{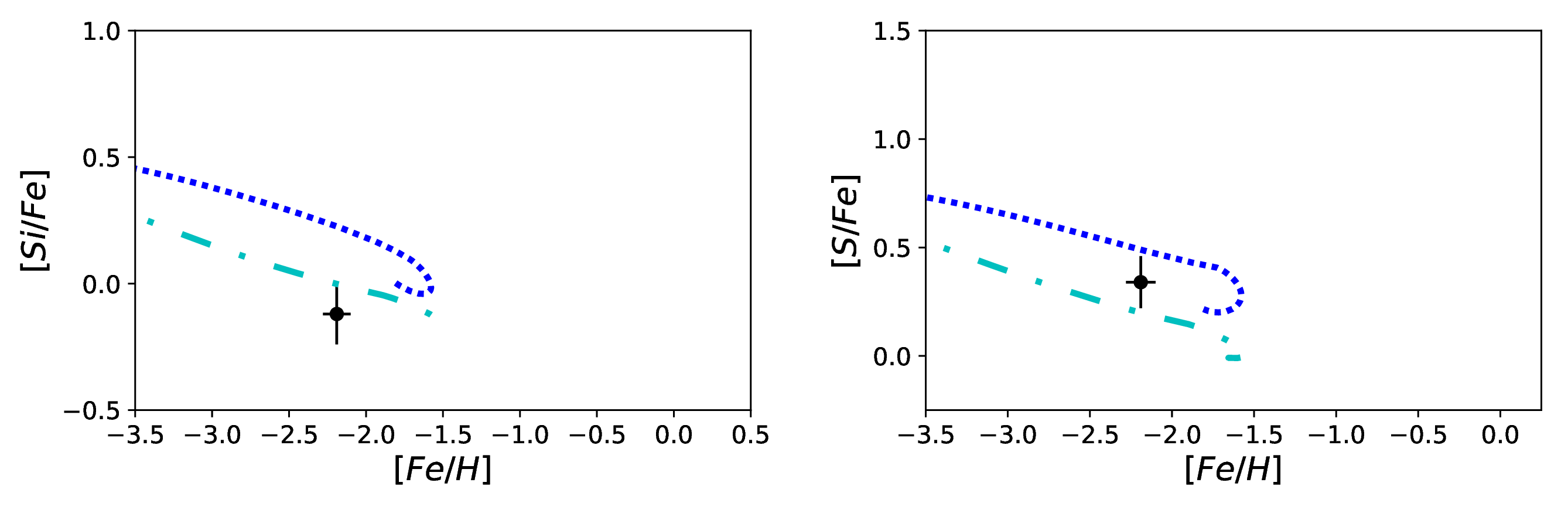}
\caption{Observed $[X/Fe]$ vs. $[Fe/H]$ ratios for the GRB 090926A host galaxy. The blue dotted (I$l$) and cyan dash-dotted line (IS$l$) are the predictions computed by means of the model for an irregular galaxy with decresed mass and SFE and the model for an irregular with decreased mass, SFE and a \citet{Scalo86} IMF, respectively.}
\label{f:Scalo_irr}
\end{figure*}
\begin{figure*}
\centering
\includegraphics[width=.81\textwidth]{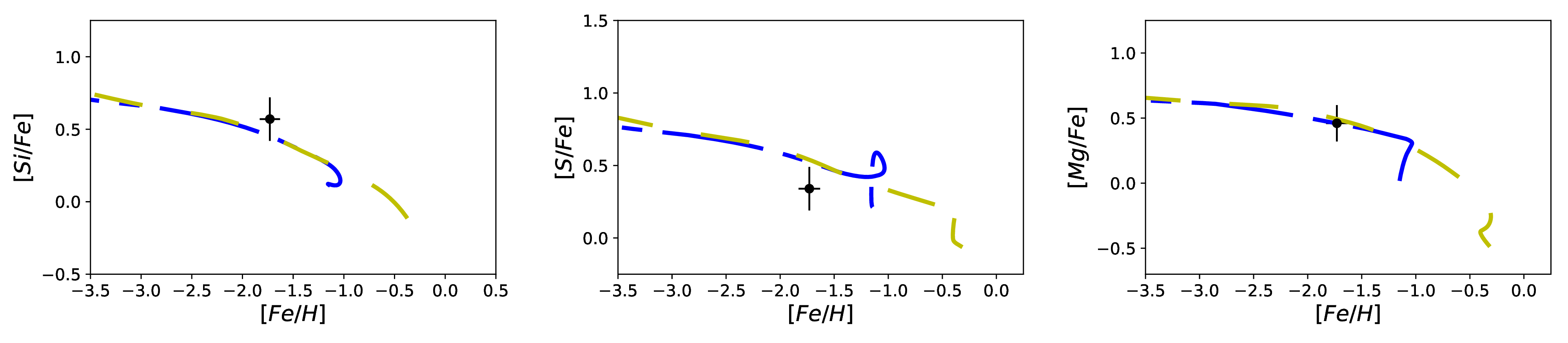}
\caption{Observed $[X/Fe]$ vs. $[Fe/H]$ ratios for the GRB 120327A host galaxy. The blue long-short dashed (I$-$) and yellow dashed line (SpS) are the predictions computed by means of the model for an irregular galaxy with decreased dust production from CC-SNe and the model for a spiral disk with a \citet{Scalo86} IMF, respectively.}
\label{f:Scalo_sp}
\end{figure*}

\subsection{IMF effects}
\label{ss:IMF_effects}
We also tried to see what happens in changing the IMF in the models for different galaxy types. The modification of this variable in fact alter significantly the results given by chemical and dust evolution equations. As mentioned in Section \ref{s:chem_model}, computations were also done adopting a \citet{Scalo86} IMF for spiral and irregular models and a top-heavy single slope IMF for the elliptical models.\\
The effects of an IMF changing are usually very effective in the abundance patterns. As an example, we see now what happens for $\alpha$-elements. Adopting an IMF that favours the formation of massive stars, we will see an overabundance in the $[\alpha/Fe]$ ratios. This is simply due to the fact that $\alpha$-elements are mainly produced by this latter class of stars. On the contrary, an IMF that disfavours the presence of massive stars will lower the $[\alpha/Fe]$ ratio. At the same time, the IMF has also an effect on the galactic winds, since these latter are also driven by CC-SN explosions.\\
However, in this work we wanted to test if a change in the IMF can give a help in identifying GRB host galaxies. We found interenting the fact that data from both the identified elliptical hosts (GRB 050820, GRB 120815A) are better explained by a model for a massive, strongly star forming elliptical galaxy ($M_{inf}=10^{12}M_\odot$, $\nu=25$ $Gyr^{-1}$) with the top-heavy IMF defined in Section \ref{s:chem_model}. In Figure \ref{f:top_heavy} is shown what happens in the case of using this latter IMF or the \citet{Salpeter55} one. From the Figure, it is evident that adopting this alternative IMF we have results much more consistent with data.
However, this result has not to surprise us. Indeed, it is coherent with previous studies on elliptical galaxies (\citealt{Arimoto86}; \citealt{Gibson97}; \citealt{Weidner13}), that claim the adoption of an IMF flatter than the \citet{Salpeter55} one to explain many of their features. Regarding to the galactic ages of the hosts, they remain of the same order of magnitude ($\sim 10$ $Myr$) of what found before.\\
In non-elliptical host galaxies, instead, we saw that in some cases a \citet{Scalo86} IMF can be more suitable to describe the obsevational data. In particular, the adoption of such IMF in the irregular models for GRB 090926A was resulted the best way to explain the very low $[\alpha/Fe]$ ratios observed. This can be seen in Figure \ref{f:Scalo_irr}. 
It remains evident the different behaviour of $S$ and $Si$ with respect to the models. As previously explained, this can be seen in terms of dust content in the galaxy. Regarding the host age, we found it only little higher with respect to the result obtained with the \citet{Salpeter55} IMF. Due to the relatively old age of this host, in fact, at that time galactic winds can play a role in the abundance patterns, compensating the slower chemical enrichement.\\
For what concerns the identified $\lq\lq$less dusty" irregular GRB 050730 and GRB 120327A hosts, we found that a spiral models with a \citet{Scalo86} IMF can adapt well to the observed abundances. Performing our statistical test, indeed, we found very similar data-model mean distances. In Figure \ref{f:Scalo_sp} is shown what we found fot GRB 120327A.
The similarity between the patterns is not surprising, since an IMF that disfavours massive star formation (as the \citealt{Scalo86} does with respect to the \citealt{Salpeter55}) will lower the $[\alpha/Fe]$. In other words, using a steeper power law for the IMF is equivalent to lower the star formation, from the point of view of the abundance patterns. Moreover, the adoption of a \citet{Scalo86} IMF for spiral galaxies is more indicated to be used. In fact, this IMF describes better the features of the MW disk than the \citet{Salpeter55} (\citealt{Chiappini01}; \citealt{Romano05}). At the contrary, for irregular galaxies it was shown that a \citet{Salpeter55} IMF is to be preferrred over a \citet{Scalo86} IMF because the first explains better many of the irregular features (\citealt{Bradamante98}; \citealt{Yin11}). We have to be aware of this latter fact in adopting a \citet{Scalo86} IMF in an irregular galaxy model, as happened for GRB 090926A.

\section{Conclusions}
\label{s:conclusion}
In this work, we present and adopt a method based on detailed chemical evolution models to constrain the nature and the age of LGRB host galaxies. This method was already used in the works of \citet{Calura09} and \citet{Grieco14} and consists in the comparison of the abundance ratios observed in GRB afterglow spectra with abundances predicted for galaxies of different morphological type (irregular, spiral, elliptical). These are obtained by means of chemical evolution models calibrated on the features of local galaxies (in the case of the irregular and spiral models) and high redshift galaxies (in the case of the elliptical models). The elements considered in this study, are $N$, $\alpha$-elements ($Mg$, $Si$, $S$) and $Fe$-peak elements ($Fe$, $Ni$ and $Zn$). For the fact that chemical abundances measured from afterglow spectra are just ISM gas abundances, the chemical evolution models take into account the dust depletion in the ISM. Concerning this latter fact, we adopt in this paper updated and more accurate prescriptions with respect to the ones used in previous works, as well as for the stellar yields. Thanks to these improvements, this paper can provide more robust insights on the nature of LGRBs and also on the star formation process in the early universe. As a matter of fact, long GRBs are supposed to be the product of the collapse of massive stars, and for this reason they can be considered as tracers of the star formation. We analysed the environment of the following 7 GRBs: GRB 050730, GRB 050820, GRB 081008, GRB 090926A, GRB 120327A, GRB 120815A, GRB 161023A.\\
We summarise our results as follows:
\begin{enumerate}
    \item The model-data comparison shows that all the three galactic morphological types (irregular, spiral, elliptical) are present in our sample of host galaxies, confirming the result obtained by \citet{Grieco14} of having also early-type galaxies as hosts of GRBs. The possibility of having massive star forming elliptical galaxies as GRB host galaxies is motivated by the high redshift of the hosts analysed in our sample ($z\apprge2$). The high values of $z$ indeed allows us to see early-type galaxies during their period of active star formation, when the massive star explosions are present. Such a situation is not present instead in the local universe, where the star formation is already quenched for this morphological type. Always concerning the elliptical galaxies, their presence in our host galaxy sample is not in contradiction with recent results (e.g. \citealt{Vergani17}; \citealt{Perley16b}).
    \item Except for the GRB 090926A, for which the predicted age is higher than $1$ $Gyr$, our models predict ages much younger than a billion year. These very short timescales are also in agreement with the low gas metallicities measured from the considered spectra. As a matter of fact, with the exception of GRB 081008, the $[Fe/H]$ values are always $< - 1.5$ $dex$. These metallicity values are compatible with the most accredited models explaining the origin of LGRBs, where the progenitors of these events have to be placed in a low metallicity environment to give rise to the phenomenon.
    \item For three of the hosts studied (GRB 081008, GRB 090926A, GRB 161023A) data seem to indicate that we are looking at more dusty environments with respect to the reference models, whose dust parameters are calibrated on observations (see \citealt{Gioannini17b}). We reached indeed a good agreement between models and data by adopting an increased dust production for CC-SNe. This indication of a large dust presence, confirms last years IR and radio band afterglow observations,  which clearly show the presence of medium to large amounts of dust in many GRB host galaxies (e.g. \citealt{Perley09}, \citeyear{Perley17}; \citealt{Greiner11}; \citealt{Hatsukade12}; \citealt{Hunt14}).
	\item Concerning $Zn$, we did not find good agreement between the observed abundances and the models: this is somewhat expected since the stellar yields for this partly s-process element are still very uncertain. Except for the $[Zn/Fe]$ found in GRB 081008, the metallicities (and consequently the galactic ages) of the host galaxies analysed are too low (short) to explain the very high $[Zn/Fe]$ values in terms of accretion of $Fe$ dust. On the other hand, higher dust production by stars (even only for $Fe$) seems to be not the solution. Nevertheless, we find an interesting $\alpha$-$Zn$ correlation, satisfied by all the hosts for which we have the availability of the $Zn$ abundances. As a matter of fact, we have very high $[Zn/Fe]$ ($>1dex$) in the case of high $[\alpha/Fe]$ , whereas in correspondence of lower $[\alpha/Fe]$ ratios typical of late-type galaxies we find lower $[Zn/Fe]$ ($\sim 0.5$-$0.6dex$). In other words, $Zn$ abundance behaves like $\alpha$-abundances.
    \item Changing the IMF in the models explains better the observed abundance ratios in some GRB hosts. In  identified ellipticals (GRB 050820, GRB 120815A) $[\alpha/Fe]$ abundances indicate the presence of a top-heavy IMF rather than the \citet{Salpeter55} one. This is in agreement with many previous results (e.g. \citealt{Arimoto86}; \citealt{Gibson97}) adopting such IMFs to explain some of the elliptical features. 
For what concerns some identified late type hosts (GRB 050730, GRB 090926A, GRB 120327A), the observed abundance ratios can be explained in terms of a \citet{Scalo86} IMF. This agrees with the studies carried out for spiral disks (\citealt{Chiappini01}; \citealt{Romano05}), but it does not for the irregulars, where the \citeauthor{Salpeter55} IMF well reproduces many of their features (\citealt{Bradamante98}; \citealt{Yin11}).    
    \item We do not always find agreement with the identification results given by \citet{Calura09} and \citet{Grieco14}. In particular, very different is the result for what concerns the GRB 050820, which host galaxy was identified by \citeauthor{Calura09} as an irregular galaxy with SFE $\nu=0.1 Gyr^{-1}$.  In this work we classify this host as a strong star forming ($\nu\ge 15 Gyr^{-1}$) elliptical. Significant differences, insofar as not strong as for GRB 050820, were also found GRB 081008 and GRB 120327A. These results strongly highlight the importance of adopting detailed new chemical evolution models with updated and more accurate prescriptions on yields and dust, as it has been done in this paper.
\end{enumerate}

\section*{Acknowledgements}
MP, FM acknowledge financial support from the University of Trieste (FRA2016). FC acknowledges funding from the INAF PRIN-SKA2017 program 1.05.01.88.04.

\bibliography{GRB_hosts}
\bibliographystyle{mnras}


\appendix


\bsp	
\label{lastpage}
\newpage

\end{document}